\documentclass[preprintnumbers,amsmath,amssymb,superscriptaddress,twocolumn]{revtex4}
\usepackage{graphicx}
\usepackage{multirow}
\usepackage{xcolor}

\usepackage{amsmath}
\usepackage{latexsym}
\usepackage{epstopdf}
\usepackage{amsmath}
\usepackage{longtable}
\usepackage{amssymb,amsmath}
\usepackage{dcolumn}
\usepackage{bm}
\setcitestyle{super}
\begin{document}

\title{Electronic, dielectric and optical properties of two dimensional and bulk ice: a multi-scale simulation study}
\author{S. Ghasemi$^{1}$,  M. Alihosseini$^{2}$, F. Peymanirad$^{3}$, H. Jalali$^{2}$, S. A. Ketabi$^{1}$, F. Khoeini$^{2}$ and M. Neek-Amal$^{3,4*}$ \\
\small$^1$School of Physics, Damghan University, P.O. Box 36716-41167, Damghan, Iran\\
\small$^2$Department of Physics, University of Zanjan, 45195-313, Zanjan, Iran.\\
\small$^3$Department of Physics, Shahid Rajaee  University, 16875-163, Lavizan, Tehran, Iran.\\
\small$^4$Department of Physics, Universiteit Antwerpen, Groenenborgerlaan 171, B-2020 Antwerpen, Belgium.
}
\begin{abstract}
 The intercalated water into nanopores exhibits anomalous  properties such as ultralow dielectric constant.~Multi-scale modeling and simulations are used to investigate the dielectric properties of various crystalline two-dimensional ices and bulk ices. Although, the structural properties of two-dimensional (2D-) ices have been extensively studied, much less is known about their electronic and optical properties. First, by using density functional theory (DFT) and density functional perturbation theory (DFPT), we calculate the key electronic, optical and  dielectric properties of 2D-ices. Performing DFPT calculations, both the ionic and electronic contributions of the dielectric constant are computed. The in-plane electronic dielectric constant is found to be larger than the out-of-plane dielectric constant for all the studied 2D-ices. The in-plane dielectric constant of the electronic response ($\varepsilon_{el}$) is found to be isotropic for all the studied ices. Secondly, we determined the dipolar dielectric constant of 2D-ices using molecular dynamics simulations (MDS) at finite temperature. The total out-of-plane dielectric constant is found to be larger than 2 for all the studied 2D-ices. Within the framework of the random-phase approximation (RPA), the absorption energy ranges for 2D-ices are found to be in the ultraviolet spectra. For the comparison purposes, we also elucidate the electronic, dielectric and optical properties of four crystalline ices (ice VIII, ice XI, ice Ic and ice Ih) and bulk water.

\end{abstract}
\maketitle

\section{Introduction}
The phase behavior of two-dimensional (2D-) ice is the subject of recent experimental and
theoretical interest, which is still controversial~\cite{nat2015,prl,prb,ACS}.
 Although, recently the report on the observation of monolayer, bilayer and trilayer ice using transmission electron microscopy (TEM)~\cite{nat2015} was challenged later~\cite{commentnat2015}, however, several theoretical studies based on both classical force fields and ab-initio simulations revealed the exciting possibility of exploring 2D-ice structures at specific conditions~\cite{prl,prb,mario,zangi}.
In particular, both classical force fields and ab-initio simulations predict that water molecules form ordered flat square lattice (f-SQ) structure while they are trapped in a few angstrom size slit~\cite{prl,mario}. The confinement width needed for the formation of stable monolayer ice is approximately h$\simeq$5-7\AA~~\cite{nat2015,zangi,mario}.

 In contrast, the structural properties of 2D-ice (and bulk ice), the electronic, dielectric and optical properties of 2D-ice and nanoconfined water is not well understood. Therefore, a solid theoretical background for the effects of size reduction on the dielectric properties of ice and confined water is highly demanded. This can be helpful to understand the measured anomalous dielectric properties of confined water~\cite{science}: recently, by using electrostatic force detection of atomic force microscope (AFM), unexpected variation in the out-of-plane dielectric constant of confined water between graphene and hexagonal boron nitride (h-BN) has been observed~\cite{science}. The presence of an interfacial water layer (having ice phase) with vanishingly small polarization is the
reason for such small out-of-plane dielectric constant ($\simeq$2) for channels with size of $h<$15\AA.~Indeed the dielectric constant of nanoconfined water was found to be about 2, which is above the high frequencies dielectric constant of water, i.e. 1.8.

In the past few decades, molecular dynamics simulations (MDS) and monte carlo simulations (MCS) have been used to calculate the dipolar dielectric constant of water and confined water.
 For instance the variation of the dipolar dielectric constant with temperature
and pressure for the ices Ih, III, V, VI, and VII was studied by Aragones et. al.~\cite{aragon}. In addition to MDS and MCS methods, mean field theory (such as Kirkwood's theory) was also used and
yielded valuable insights into the H-bonding effects on water dielectric constant~\cite{daniel}. Notwithstanding the existing MD based theory studies in the past few years, the questions about the ionic and electronic contributions of dielectric constant of 2D-ice is still unanswered. Most of the previous studies reported the dipolar dielectric constant of water confined at the nanoscale channels~\cite{ref7,ref8,ito}. Thus, one naturally expects to quantify the dipolar, electronic and ionic dielectric constants of crystalline 2D-ice and corresponding frequency dependence. This will provide a solid theoretical support for the recent experiment~\cite{science}.

Here, we conducted a systematic study for detrmining the electronic, dielectric and optical properties of 2D-ice using multi-scale approach including first principles and molecular dynamics simulations. Both the out-of-plane and in-plane components of ionic, electronic and dipolar dielectric constants of stable structure of 2D-ices including flat square (f-SQ), buckled square (b-SQ), buckled rhombic (b-RH) and hexagonal structure (HEX) are investigated~\cite{prl}. We also reported results for the electronic, dielectric and optical properties of the crystalline bulk ice and bulk water.
Our work provides benchmark theoretical data for the electronic, dielectric and optical properties of crystalline 2D- and bulk ices.

\begin{figure*}[]
\includegraphics[width=.08\linewidth]{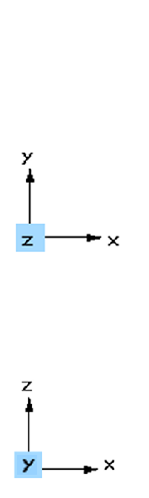}
\includegraphics[width=.220\linewidth]{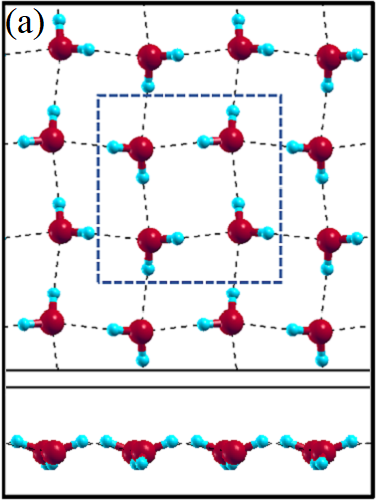}
\includegraphics[width=.220\linewidth]{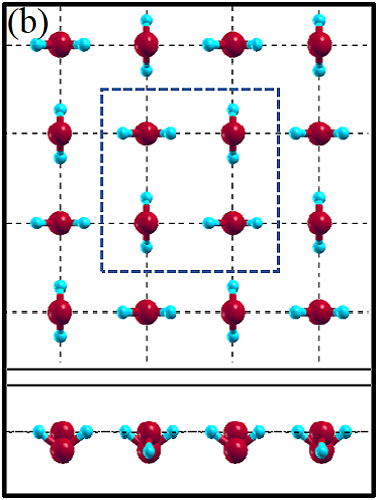}
\includegraphics[width=.220\linewidth]{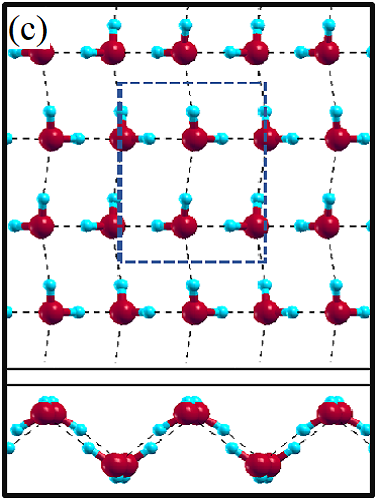}
\includegraphics[width=.220\linewidth]{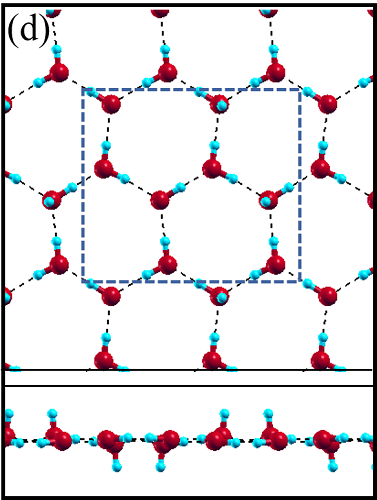}
\includegraphics[width=.08\linewidth]{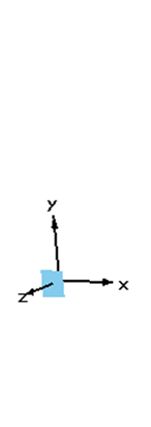}
\includegraphics[width=.220\linewidth]{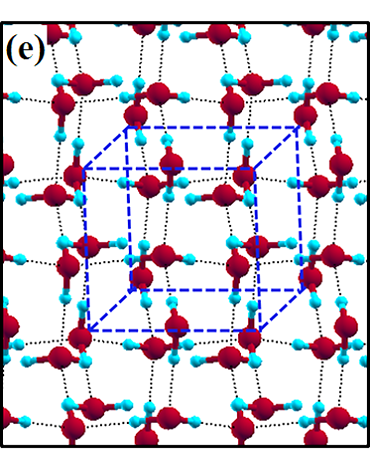}
\includegraphics[width=.220\linewidth]{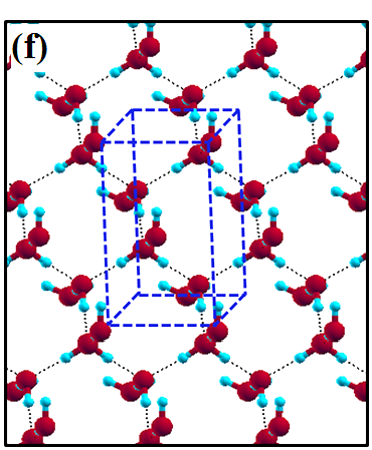}
\includegraphics[width=.220\linewidth]{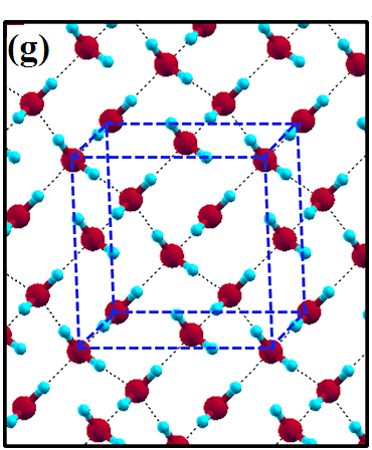}
\includegraphics[width=.220\linewidth]{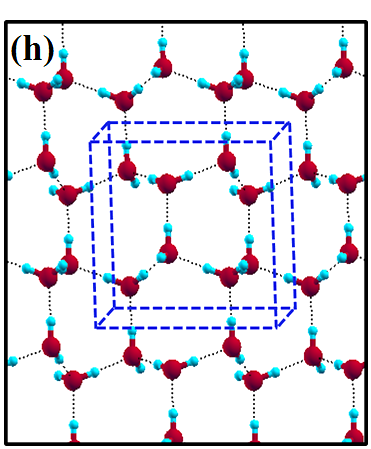}
\caption{ The lattice structure of two-dimensional and bulk ices. The top and side views of (a) flat square (f-SQ), (b)
 buckled square (b-SQ), (c)  buckled rhombic (b-RH), and (d) hexagonal (HEX) 2D-ice structures. The blue squares show the
 corresponding supercells. (e) ice VIII, (f) ice XI, (g) ice Ic, and (h) ice Ih. The blue cubes show the corresponding unitcells.} \label{fig1}
\end{figure*}

\section{Two-dimensional ice}

The 2D-ice structures considered in this paper are flat square (f-SQ), buckled square (b-SQ), buckled rhombic  (b-RH) and hexagonal (HEX). Their optimized structures were reported in Ref.~[2]. The f-SQ and b-RH (HEX) structures have square (rectangular) unitcell containing 4 (8) water molecules (see Figs.~\ref{fig1}(a-d)). In order to eliminate the interaction between periodic images in ab-initio calculations, we set  a large vacuum with height ``$c$" for calculating each 2D-ice (see section ``Dielectric constant"). The lattice parameters of the simulation supercells ($a,~b,~c$) are listed in Table 1. The input  lattice structures were extracted from the optimized structures of Ref.~[2]. Using van der Waals diameter of oxygen atom, one can approximate the effective thickness ($t$) of a given 2D-ice lattice: considering flat structure of  f-SQ, $t_{f-SQ}=t_0=2\times R_v$, where $R_v$ is the van der Waals radius of oxygen atom. For b-SQ and HEX structures, there is  1 angstrom difference between top row and bottom row of O atoms (see side view of b-SQ and HEX structures in Fig. 1(b,d)), thus $t_{b-SQ,HEX}=1+t_0$. For the b-RH, there is a large (4\AA)~ distance between the top row and bottom row of O atoms, i.e $t_{b-b-RH}=4+t_0$.  All the structure parameters of the studied systems and the important findings of this papaer are tabulated in Table 1.

\section{Crystalline bulk ice}
In order to compare electronic, dielectric and optical properties of above mentioned 2D-ices with the high pressure phase of ice, we also calculated the dielectric constant of four crystalline ice, i.e. cubic (ice VIII and ice Ic) and hexagonal (ice XI and ice Ih) bulk ices.\\
The ice XI, ice Ic and ice Ih structures have 8 water molecules per orthorhombic unitcell (see Figs.~\ref{fig1}(e-h)). The lattice parameters of the crystalline ices are listed in Table I.  The ice Ih has the hexagonal crystalline form of ordinary ice, which is stable at temperature 273\,K (down to few Kelvin) and pressures up to 200\,MPas. The ice Ic is one of the metastable cubic crystalline form of ice, which is stable at temperatures between 130 and 220 K. The ice VIII with 8 water molecules per unitcell has a tetragonal crystalline form and is stable under high pressures about 3\,GPa below 278\,K. The ice XI is a hydrogen-ordered form of ice Ih containing 8 molecules per unitcell and is stable at temperature 5\,K and  pressures of about 100\,MPas.
We also calculated the dielectric constant of bulk water using 17 water molecules inside a cubic unitcell with size 7.93$\times$7.93$\times$7.93\,\AA$^3$. We have used 20 different relaxed MDS configurations as inputs for the density functional theory (DFT) calculations of bulk water. For the bulk water and bulk crystalline ices, we applied periodic boundary conditions in all directions, though for the 2D-ices, a vacuum space must be set (see section V. A). In order to obtain more insights and for comparison purposes, we put in order the dielectric constant of some common 2D-materials such as monolayer h-BN and monolayer of three transition metal dichalcogenides (TMDs)~\cite{2Dmater}, i.e. MoS$_{2}$, WS$_{2}$, and WSe$_{2}$.

\section{Dielectric constant}

For the polar systems, the total dielectric constant tensor includes three main contributions, i.e. electronic, ionic and dipolar:
\begin{equation}\label{Eq1}
\varepsilon_{total}^{\mu\nu}=\varepsilon_{el}^{\mu\nu}+\varepsilon_{ion}^{\mu\nu}+\varepsilon_{dip}^{\mu\nu},
\end{equation}
The indexes $\mu$ and  $\nu$ run over the three spatial directions. At zero Kelvin the dipolar term vanishes. By increasing temperature of 3D-material, the dipolar term becomes important and should be taken into account. Note that different bulk ices are stable at diffrent tempratures~\cite{Martin}. In Table I, the relevant temperatures are listed.

\section{The methods}
In this study, density functional theory (DFT) has been implemented for electronic band structure calculations. We have used density functional perturbation theory (DFPT) to obtain the electronic and ionic dielectric constants at zero Kelvin. Moreover, molecular dynamics simulations (MDS) has been used to find the dipolar dielectric constant at a finite temperature for 2D-(bulk) ices. In order to calculate the optical dielectric function, the random-phase approximation (RPA)~\cite{Brener1975} based on DFT ground-state calculations has been conducted. In the following sections, we briefly explain the different used methods.

\subsection{Density functional theory: Electronic band structure}
We have calculated the electronic band structure of 2D-ices using
density functional theory (DFT) as implemented in the Quantum-ESPRESSO (QE) package~\cite{Giannozzi2009}. We have used ultrasoft pseudopotentials to treat the interaction between the ion cores and valence electrons and applied the generalized gradient approximation (GGA) for the exchange-correlation interactions. We also have studied the effect of nonlocal correlations using the van der Waals density functional (vdW-DF) of Dion et al~\cite{Dion2011}. Accuracy of forces on each atom has been considered about 0.1 mRy/bohr for the variable-cell optimization, relaxing the cell parameters and atomic positions. in order to accurate calculation of structural and electronic properties, the kinetic energy cut-offs of $150~Ry$  and $1500~Ry$ were found to be sufficient for the wavefunctions and the charge densities, respectively, where the  k-point grid for 2D- (bulk) ices was set to $6\times6\times1$ ($6\times6\times6$). The k-point grid for 2D- (bulk) ices was chose $24\times24\times1$ ($24\times24\times24$) for non-self-consistent calculation in partial density of state (PDOS) analysis. The smearing parameter of 0.01 Ry has been used for PDOS analysis. In addition, total energy convergence threshold was set to $10^{-12}eV$. The Coulomb cutoff technique~\cite{Rozzi2006,Sohier2017} was used to reduce interactions between periodic images and cost of the ab-initio calculations for 2D-ice structures.

\subsection{Density functional perturbation theory: Electronic and ionic dielectric constant}
The dielectric properties of the 2D-ices were determined using the DFPT approach. In DFPT, the dielectric constant tensor is defined as a linear response to the perturbative electric field~\cite{baroni2001,Gnoze1997} and the ionic displacement are considered as a perturbation to the equilibrium system. The response of the electronic charge density to the perturbative electric field in the linear response regime, employed to determine the electronic contribution of the dielectric tensor, i.e. the high-frequency dielectric constant ($\varepsilon_{el}$). Subsequently, the static dielectric constant ($\varepsilon_0^{\mu\nu}$) is the summation of the electronic and the ionic parts of the system to the applied electric field:
\begin{equation}\label{Eq2}
\varepsilon_0^{\mu\nu}=\varepsilon_{el}^{\mu\nu}+\varepsilon_{ion}^{\mu\nu},
\end{equation}
where
\begin{equation}\label{Eq3}
\varepsilon_{ion}^{\mu\nu}=\frac{4\pi}{\Omega}\\\sum_{m}\frac{S_m,\mu\nu}{\omega_m^2}.
 \end{equation}
 Here $S_{m,μν}$ is the mode oscillator strength tensor, defined in terms of the Born effective charges $Z^*$ , the atomic masses $M_i$, and the normalized eigenvectors $u_{i,m\mu}$ of the i$^{th}$ ion along a given direction $\mu$ for a particular mode $m$. Also $\omega_m$ is the phonon mode frequency and $\Omega$ is the unitcell volume. Thus, in order to compute $\varepsilon_0$, the knowledge of all the phonon frequencies at the
 zone center of  Brillouin zone is needed. The latter requires the solution of the dynamical matrix at the zone center.

\begin{figure*}[!ht]
\includegraphics[width=.49\linewidth]{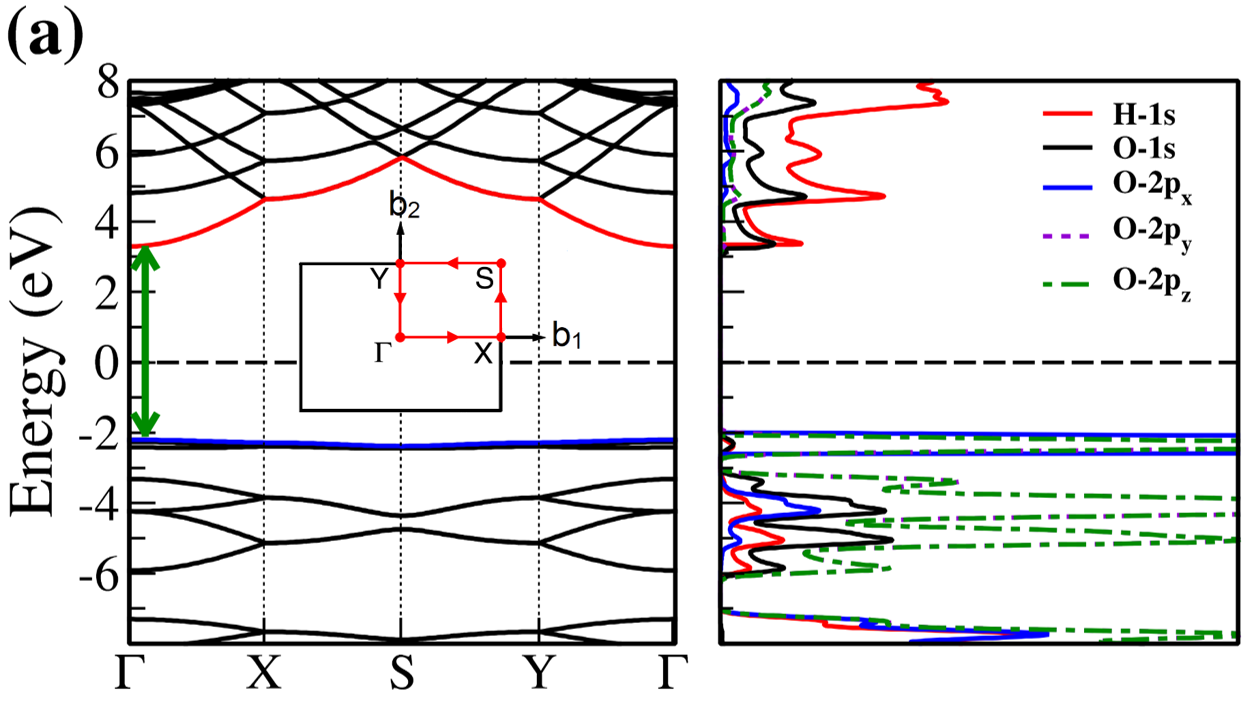}
\includegraphics[width=.49\linewidth]{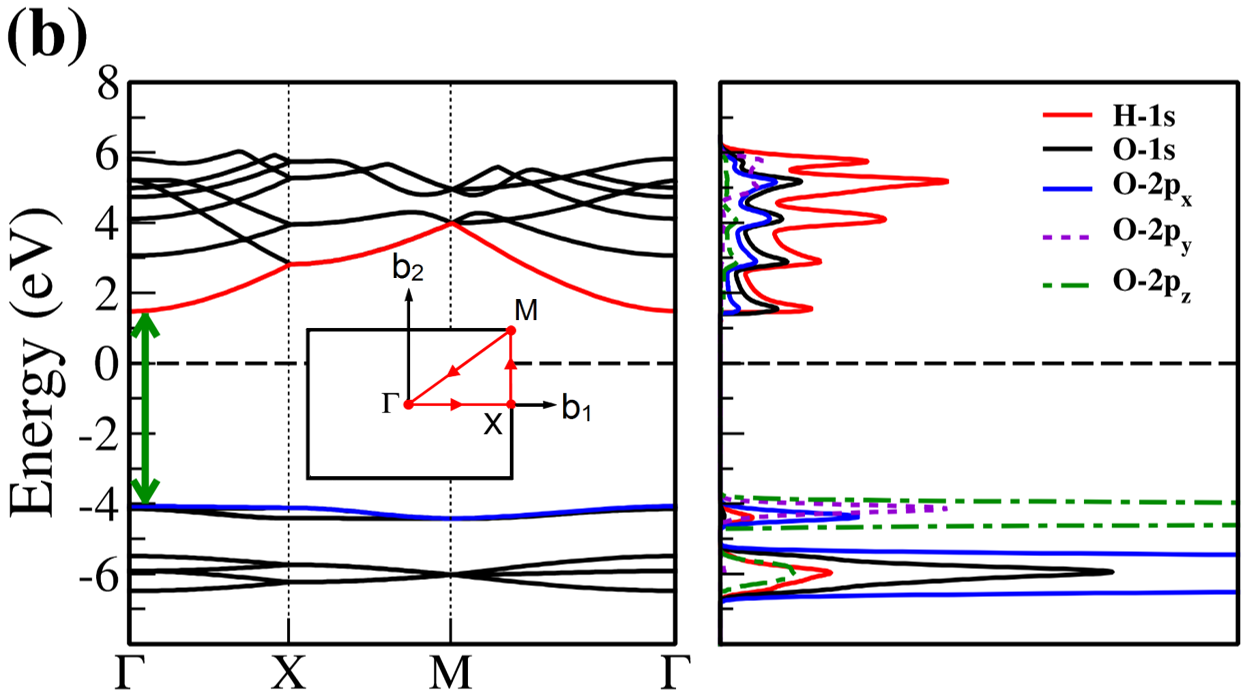}
\includegraphics[width=.49\linewidth]{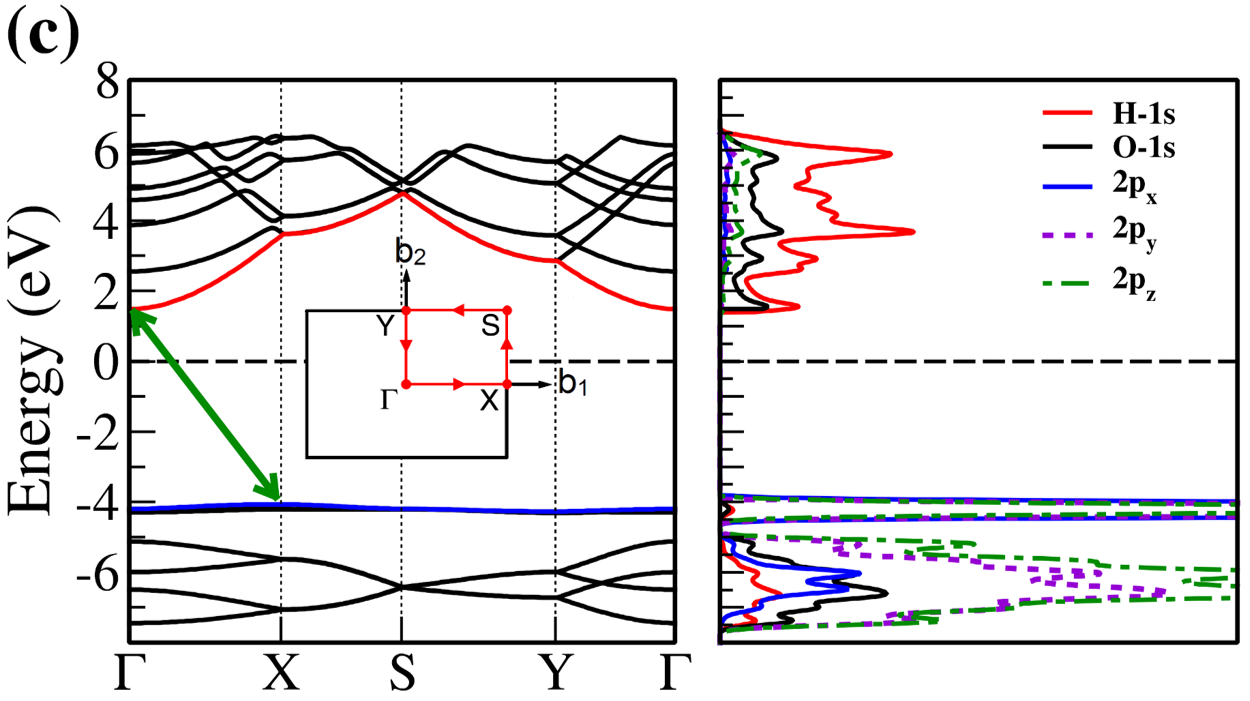}
\includegraphics[width=.49\linewidth]{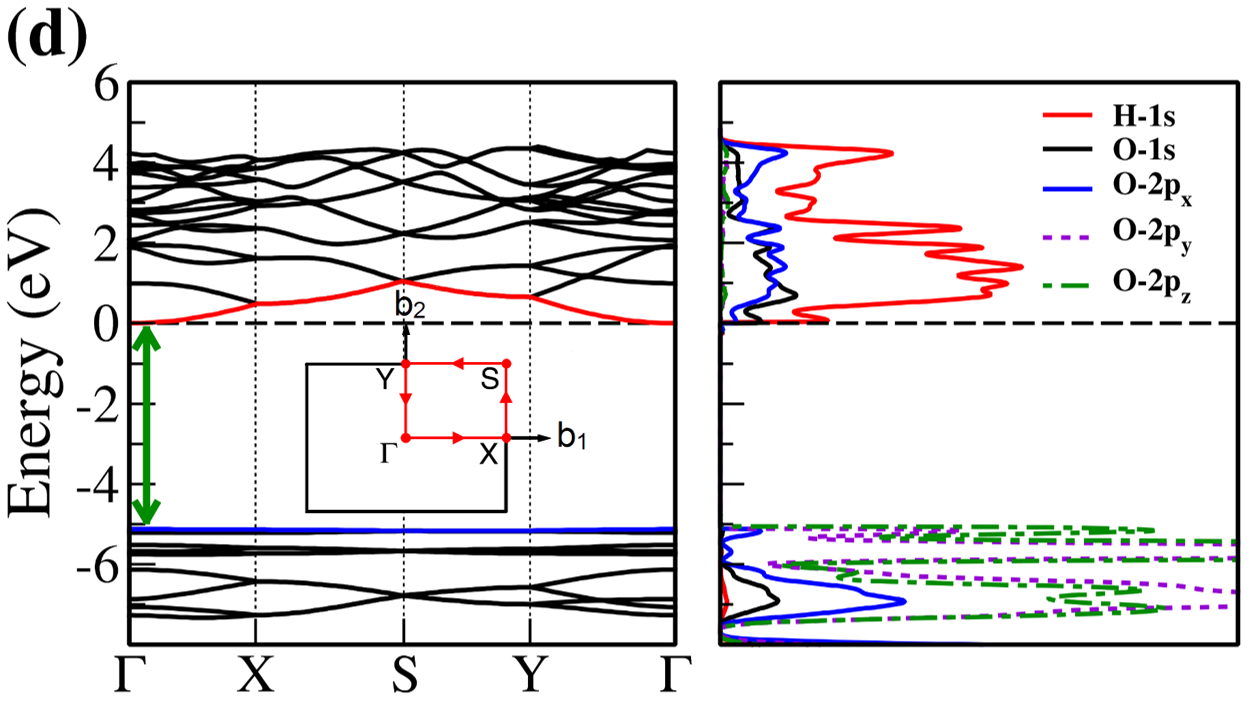}
\includegraphics[width=.49\linewidth]{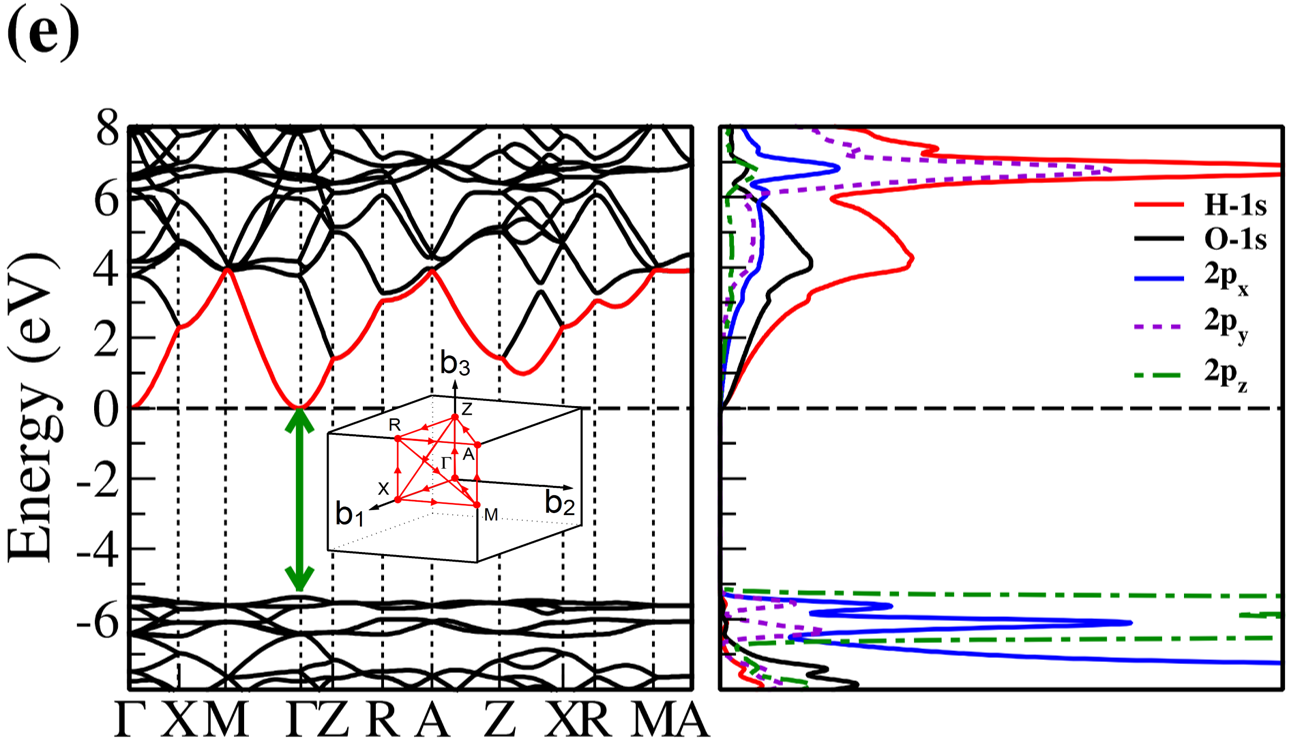}
\includegraphics[width=.49\linewidth]{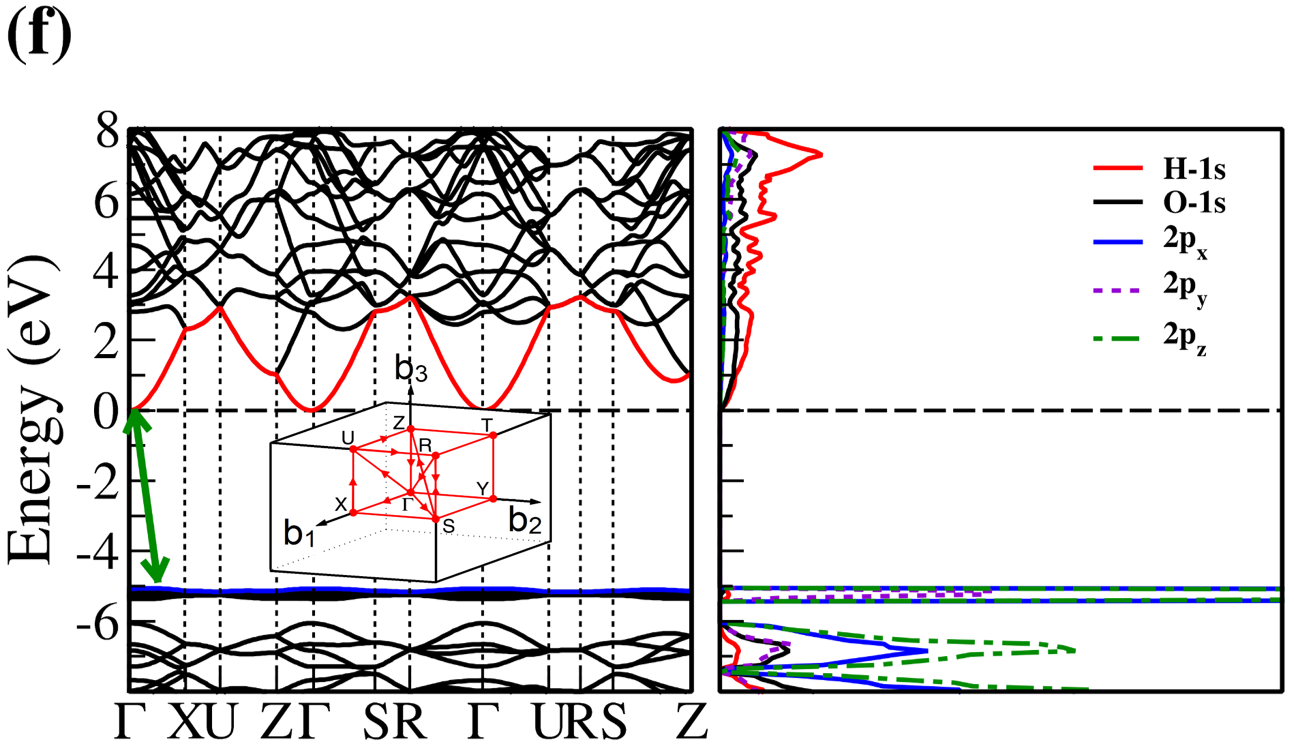}
\includegraphics[width=.49\linewidth]{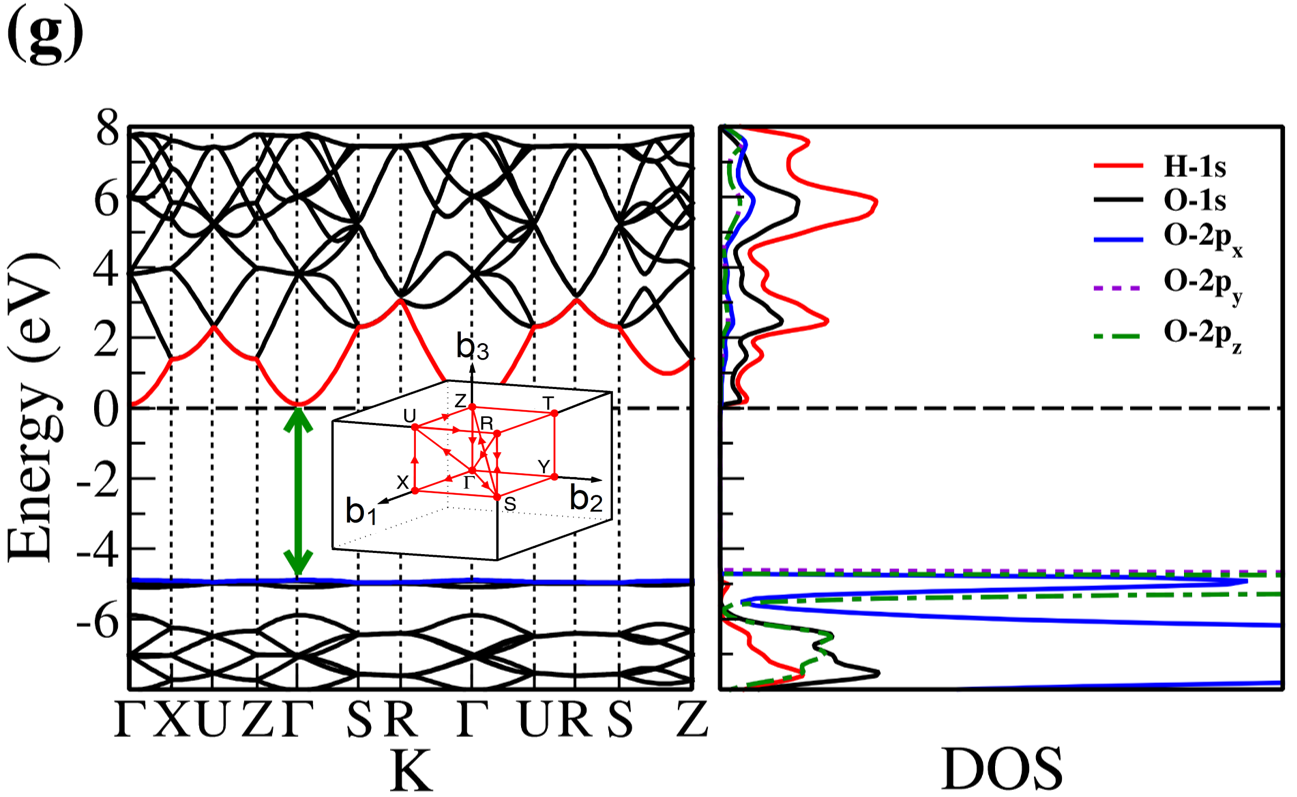}
\includegraphics[width=.49\linewidth]{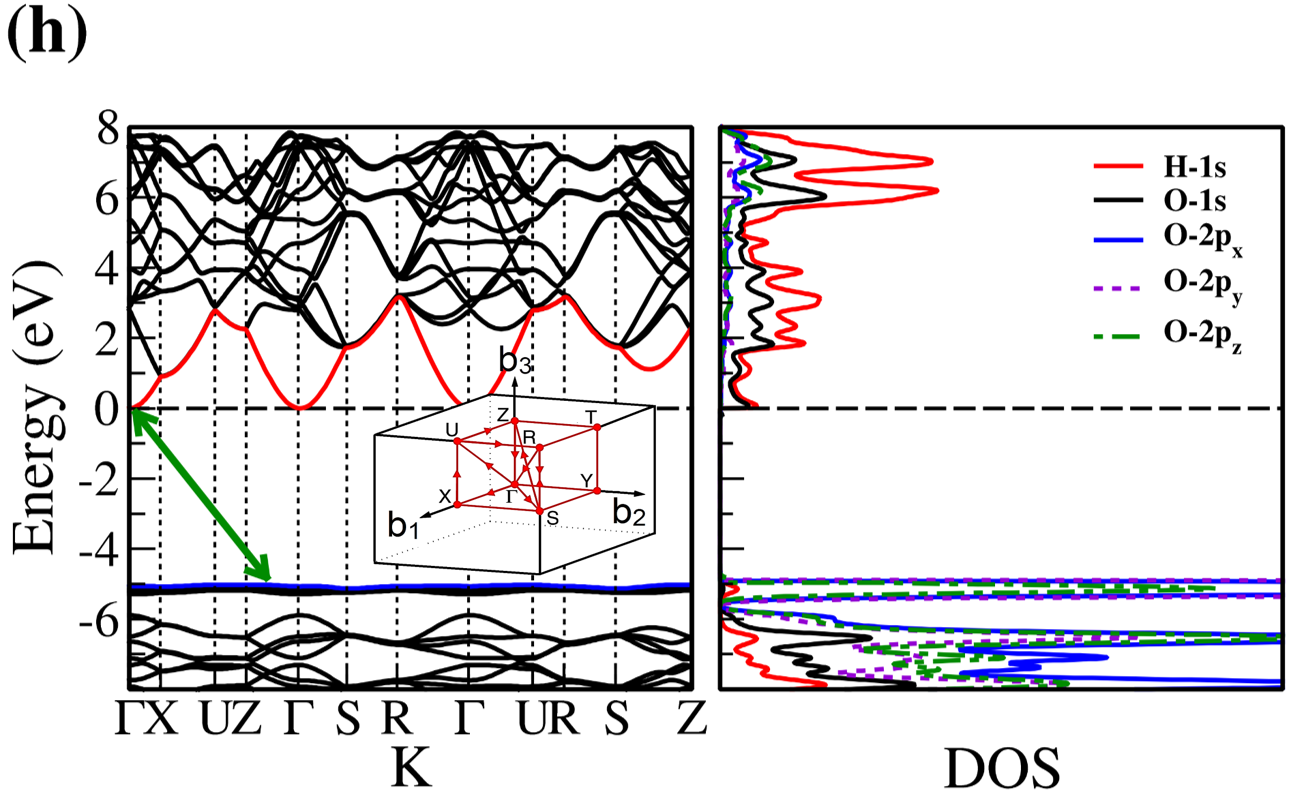}
\caption{ The electronic band structure of (a) f-SQ, (b) b-SQ, (c) b-RH, (d) HEX, (e) ice VIII, (f) ice XI, (g) ice Ic,
 and (h) ice Ih and corresponding partial density of states. The  arrows denote the type of band gap (direct or indeirect) and the insets show
 the corresponding Brillouin zones. } \label{fig2}
\end{figure*}
\subsection{Molecular dynamics simulations: Dipolar dielectric constant}

By means  of equilibrium molecular dynamics
simulations employing the large scale atomic/molecular massively
parallel simulator LAMMPS~\cite{{lammps}}, we computed the molecular (dipolar) dielectric constant of 2D-ice at 80 K. Our simulated systems contain 2D-ices, confined between two walls, separated by a distance~\cite{prl} $h=2\,t$ (these walls are used to produce the confining potential). The number of water
molecules are 400, 100 and 128 for f-SQ, R-SQ, and HEX, respectively\cite{1810}.
 The TIP4P model was employed to describe the water molecules~\cite{TIP4P}. The NVT ensemble
(Nose-Hoover thermostat) is used to keep the temperature at 80 K for 2D- and bulk ices excluding ice Ih (200 K). We modified TIP4P model to incorporate
the lattice structure of the optimized structures by using DFT: we changed bond lengths and bond angles of water molecules relevant to the DFT outputs~\cite{prl}. Also, the charges of O and H atoms were modified such that the large dipole moment ($\sim$ 3D) for single water molecule is reached~\cite{aragon}. In order to verify  this modification, we calculated the variation of dipolar dielectric constant of bulk water with time using the modified TIP4P model, see  Fig. 3(a). It is seen that the dielectric constant of bulk water lies in acceptable range, i.e. it is between dielectric constant of bulk water using  TIP4P and TIP4P/2005 models.

Periodic boundary conditions are applied along $x$, $y$ directions
and the confinement was along the $z$-direction. The
particle-particle particle-mesh method was used to compute the
long-range Coulomb interaction with a relative accuracy of
10$^{-4}$. Water bond lengths and bond angles were fixed by the SHAKE
algorithm~\cite{shake}. In all MDS a time step of 1\,fs was
chosen. Following the system relaxation (1\,ns), the thermodynamical
sampling was done up to 8\,ns to ensure the smoothness of the
converged dielectric constant.
The temperature in MDS for bulk ices (water) set to be 80 K (300 K); because all of the studied bulk ices are stable at this temperature.

A microscopic picture of dielectric properties of 2D-ices could be found by calculating the fluctuations
of the total polarization of a system, $\vec{M}$ at finite temperature. By calculating different components of the total dipole moment i.e. $M_x$, $M_y$, and $M_z$ after equilibration,  one obtains different components of the molecular dielectric constant tensor as~\cite{zhang}
\begin{equation}
\varepsilon_{dip}^{\mu\nu}=\varepsilon_{\infty}+\frac{\sigma^2_{\mu\nu}}{\varepsilon_0 k_B T
V},\label{eq1}
\end{equation}
where $\varepsilon_{\infty}$ is the optical dielectric constant and taken to be 1. Also, $\sigma_{\mu\nu}^2$ $=$ $\langle
M_{\mu}M_{\nu}\rangle-\langle M_{\mu}\rangle \langle
M_{\nu}\rangle$ while $\mu,\nu=x,y,z$ and $V$ is the volume of
the system. Here, the time averaging was taken for more than 5\,ns when in-plane dielectric constant is converged. Note that Eq. (4) can only be used
 for a homogeneous systems~\cite{ref7}.

\subsection{Random phase approximation: The frequency dependent optical  dielectric constant}
The optical dielectric function $\varepsilon_{el}(\omega)$ was  calculated using norm-conserving pseudopotentials, in the energy range of 0 to 30\,eV. In order to increase accuracy for the dielectric functrion calculations, we used the k-point grid for 2D- (bulk) ices as $12\times12\times1$ ($12\times12\times12$).
The optical dielectric constant was calculated (the frequency dependent of the electronic dielectric constant or equivalently the electronic part of the dielectric function) within the framework of the random-phase approximation (RPA)  based on DFT ground-state calculations. The mentioned dielectric function consists of frequency dependent real ($\varepsilon_{el}^r(\omega)$) and imaginary part ($\varepsilon_{el}^i(\omega)$). It is represented as:
\begin{equation}\label{Eq3}
\varepsilon_{el}(\omega)=\varepsilon_{el}^r(\omega)+i \varepsilon_{el}^i(\omega)
\end{equation}
The imaginary part of dielectric function ($\varepsilon^i(\omega)$) can be calculated using Kubo–Greenwood formalism~\cite{Morgan1988}.
Once we know the imaginary part, the real part can be obtained using the Kramers–Kronig relations (for more details see Ref.~[29]). Note that the optical dielectric constant extracted from RPA (at zero frequency) is compatible to that obtained from DFPT.

\section{Results and discussion}
\subsection{Electronic band structure}

\begin{figure}[]
\includegraphics[width=0.95\linewidth]{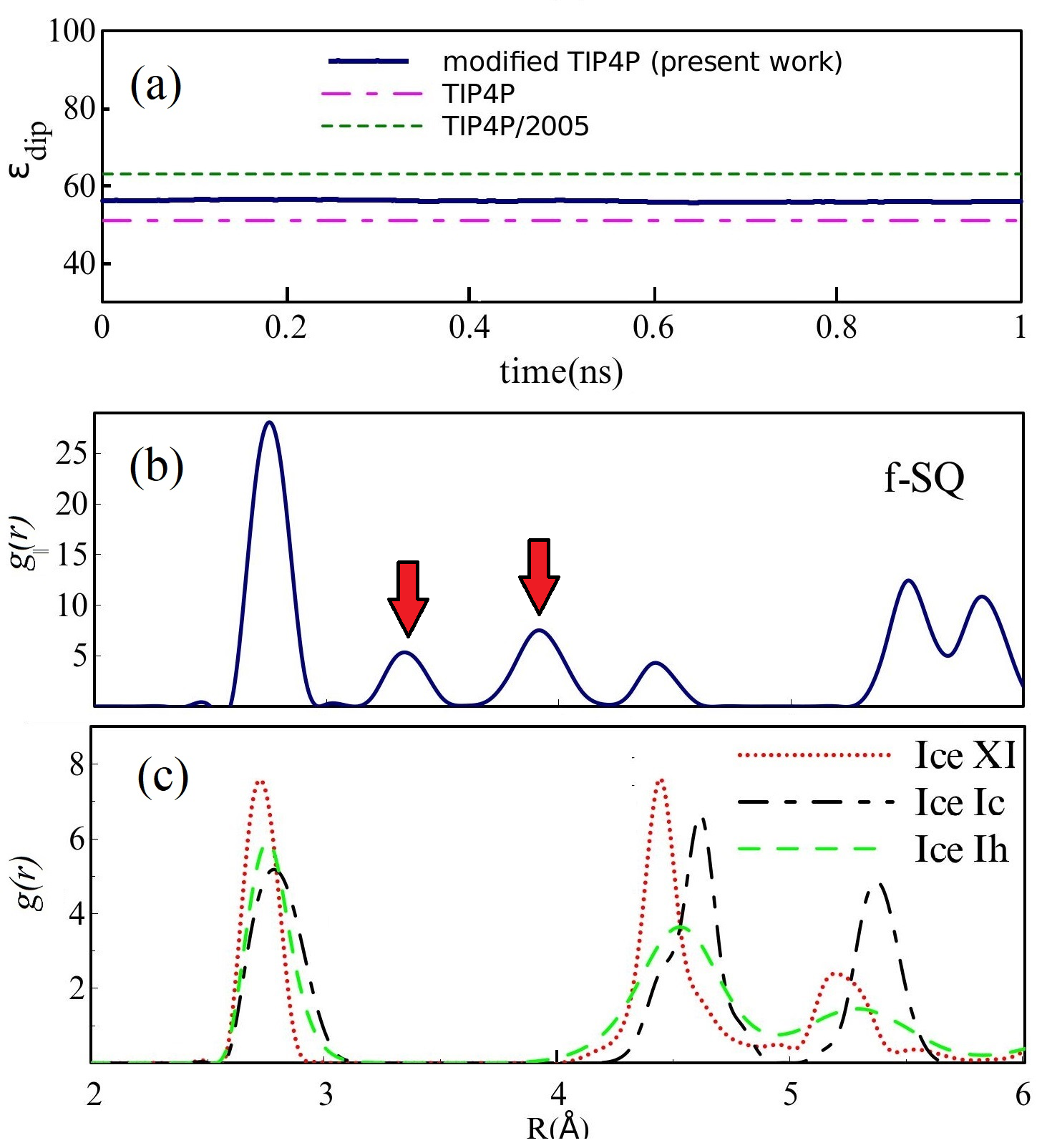}
\caption{(a) The variation of dielectric constant of bulk water with time using the modified TIP4P model in this study. (b) The in-plane radial distribution function of f-SQ 2D-ice. Two arrows indicate the emergence of two peaks in f-SQ which is absent in bulk ices. (c) The radial distribution function of three typical bulk ice. } \label{fig03}
\end{figure}

\begin{figure}[]
\includegraphics[width=0.85\linewidth]{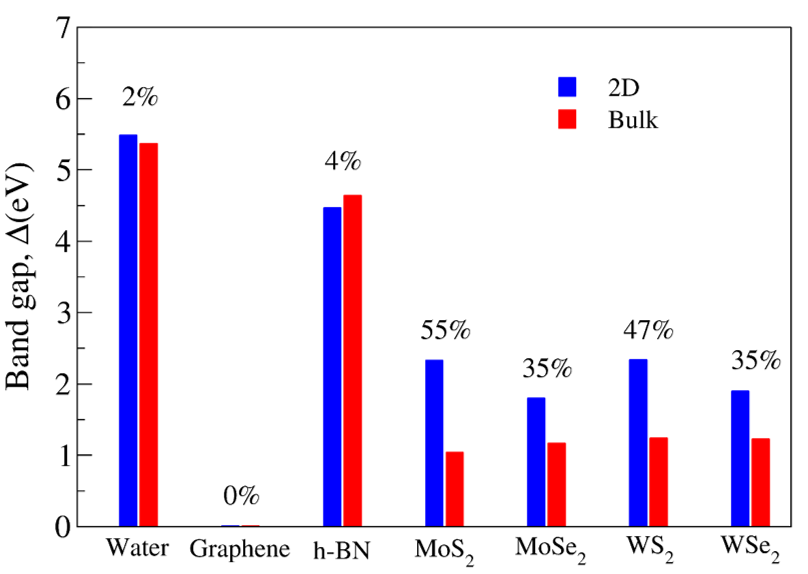}
\caption{ The energy gap of bulk phase and 2D- phase of various materials.
 The data for transition metal dichalcogenides were taken from Refs.~[30,31].} \label{fig3}
\end{figure}
\begin{table*}[!ht]
  \centering
\caption{Simulation parameters used for the calculation of
dielectric constant of 2D-ices and bulk ices: ionic dielectric constant ($\varepsilon_{ion}$), the electronic dielectric constant ($\varepsilon_{el}$), the dipolar dielectric constant ($\varepsilon_{dip}$), total dielectric constant ($\varepsilon_{tot}$)\cite{Notice}, the energy gap ($\Delta(eV)$), and the effective thickness ($t$). For comparison purposes we show the corresponding data for bulk water, TMDs and h-BN. The used temperatures in the molecular dynamics simulations are given too.}\label{table1}
\begin{center}
\begin{tabular}{c|cccccc|ccc|ccc|ccc|ccc|cc}
    \hline
    \hline

& \multirow{3}{*}{structure} & \multirow{3}{*}{a} &\multirow{3}{*}{b} & \multirow{3}{*}{c}& \multirow{3}{*}{t}& \multirow{3}{*}{$\Delta(eV)$} &\multicolumn{6}{c|}{$\varepsilon_{0}$}&& \multirow{2}{*}{$\varepsilon_{dipolar}$}&&&\multirow{2}{*}{$\varepsilon_{total}$}&& \multirow{3}{*}{T(K)}& \multirow{3}{*}{Reference}\\
\cline{8-13}
 & & & & & & & \multicolumn{3}{c|}{$\varepsilon_{el}$} & \multicolumn{3}{c|}{$\varepsilon_{ion}$}&& &&& & && \\
\cline{8-19}
 & & & & & & & xx & yy & zz& xx & yy & zz & xx & yy & zz & xx & yy & zz & &\\

  \hline

\multirow{4}{*}{\rotatebox{90}{‏2D ice}}
& f-SQ & 5.84 & 5.84 & 15 & ($t_0$=)3& 5.49& 2.0 &2.0 & 1.7& 1.0 &1.0 & 2.0 & 1.55& 1.54 & 2.41 & 3.55& 3.54& 5.11& 80 & present work\\

& b-SQ & 5.78 & 5.78 & 20 &4& 5.55& 1.8 & 1.8 & 1.49 &3.85 &3.9 & 0.01 & -&- &-& -& -& -& Ref.~[24] & present work \\

& b-RH & 4.47 & 5.76 .& 20 & 8 & 5.55 & 1.45 & 1.45 & 1.3 &10.75& 0.95 & 1.0 &1.49 & 1.52& 1.92 & 12.69& 2.92& 3.22& 80 & present work \\

& HEX & 8.7 & 7.72 & 15 &4.0 &5.12 &  1.71 & 1.71 & 1.45 & 2.0 & 1.90 & 3.53 & 3.02 & 4.32 & 2.49& 5.73& 6.93& 6.47& 80& present work  \\
\hline

\multirow{4}{*}{\rotatebox{90}{‏Bulk ice}}
& ice VIII & 4.73 & 4.73 & 6.85&- & 5.37 & 2.36 & 2.36 & 2.34 & 3.96 &3.96 & 0.26& 8.71 & 8.54 &1.03& 14.03& 13.86& 2.63& 80 &  present work \\

& ice XI & 4.45 & 7.7 & 7.27 &-& 5.06 &  1.83 & 1.82 & 1.82 & 1.81 & 0.83 & 0.93& 2.06 & 1.31& 1.86 & 4.7& 2.96& 3.61& 80 &  present work \\

& ice Ic & 6.48 & 6.48 & 6.48 &- & 4.99 & 1.77 & 1.77 & 1.76 &  1.21 & 1.21 & 0.16 & 2.1 & 2.1 & 1.01 &4.08& 4.08& 1.93& 80 &  present work \\

& ice Ih & 7.81 & 7.38 & 4.52 &- & 5.05 &  1.80 & 1.80 & 1.80 &1.14 & 1.0 & 0.96 & 1.94 & 1.92 & 1.5 & 3.88& 3.72& 3.26& 200 &  present work \\
\hline

\rotatebox{90}{\parbox{0.7cm}{‏Bulk}}& water &7.93 & 7.93 & 7.93&-& 4.99&  1.85& 1.84& 1.86& 1.28&1.16&1.29& 72&72&72& 74.13& 74& 74.15& 300 & present work \\
 \hline

& h-BN & 2.51 & 2.51 & 25.1 & 3.17 & 5.97 &  4.97 & 4.97 & 2.89 &1.85 & 1.85 & 0.4 & - & - & -  & -& -& -& & Ref.~[14] \\
\hline
\multirow{3}{*}{\rotatebox{90}{‏TMDs}}
& MoS$_{2}$ & 3.21 & 3.21 & 32.1 & 6.12 & $<2$ &   15.1 & 15.1 & 6.4 &0.2 & 0.2 & 0.0 & -& -& - & -& -& -& & Ref.~[14] \\
& WS$_{2}$ & 3.21 & 3.21 & 32.1 & 6.14 &  $<2$ &   13.6 & 13.6 & 6.3 & 0.1 & 0.1 & 0.0 & -& -& -& -& -& -& &  Ref.~[14] \\
& WSe$_{2}$ & 3.34 & 3.34 & 33.4 & 6.52 & $<2$ &  15.1 & 15.1 & 7.5 & 0.2 & 0.2 & 0.0  & -& -& -& -& -& -& &  Ref.~[14] \\
 \hline
  \hline
\end{tabular}
\end{center}
\end{table*}

The electronic band structure and partial density of states (PDOS) of 2D- and bulk ices are shown in Fig.~\ref{fig2}.
We found the direct band gap of energy about 5.49, 5.55 and 5.12 $eV$ for f-SQ, b-SQ and HEX 2D-ices and 5.37 and 4.99 $eV$ for ice VIII and ice Ic, respectively.
The electronic band structure for b-RH 2D-ice, ice XI and ice Ih confirm the indirect band gap. The corresponding energy gap for b-RH, ice XI and ice Ih are 5.55, 5.06 and 5.08 $eV$, respectively. The PDOS are shown in the right side of the panels of Fig.~\ref{fig2}. It is seen that in contrast to the conduction bands, the O (H) atoms contribution in the valance band is larger (negligible). Notice that the difference between the energy gap of 2D-ices (and 2D-h-BN) and bulk water (bulk h-BN) is small, i.e.  2$\%$ (4$\%)$,  as compared to the difference between bulk TMDs and 2D-TMDs, see Fig.~\ref{fig3}. This is due to the weak hydrogen bond between water molecules in bulk and 2D-ices as well as insulating feature in all phases of water and h-BN. The TMDs have large reductions of energy gap while transiting from 3D to 2D. The obtained energy gap, the electronic band structure and PDOS for ice VIII, ice XI,and ice Ih are in agreement with the results of Refs.~[32-35].

Independent of water phase (including ice), the nearest neighbor distance between oxygen atoms of water molecules are almost the same, i.e. 2.8-3\AA~ i.e. all the ice phases formed under very high pressures satisfy the so-called Bernal-Fowler ice rules where each water molecule has four hydrogen-bonded neighbors with a quasi-tetrahedral configuration with two short O-H distances (the donated protons) and two long ones (the accepted protons). Transiting from bulk into 2D phase of ice, the crystalline structure with larger density can be formed where the nearest neighbor distance are more or less the same but, the preferential tetrahedral bonding geometry is different.  In Figs. 3(b,c) we depict the in-plane radial distribution function for O-O distance in f-SQ and 3D-radial distribution function of three bulk ice (Ice XI, ice Ic and ice Ih). As can be seen, as expected the nearest neighbor O-O distance (first peaks) for all structures are equal. However, the second nearest neighbors (shown by arrows in Fig. 3(b)) are different. The latter causes a larger density of f-SQ~\cite{mario} (1.36gcm$^{-3}$) as compared to bulk crystalline ices i.e. $~$ 0.92 gcm$^{-3}$ ice Ih and $~$ 0.92 gcm$^{-3}$ ice XI.\\
On the other hand, the most well-known electrical conductivity mechanism for all phases of water/ice is the Grotthuss mechanism (GM) also known as proton jumping. In GM an excess proton or proton defect diffuses through the hydrogen bond network of water molecules and a covalent bond between neighboring molecules are formed and broken continuously. Because of the same nearest neighbor distance in bulk ice (bulk water) and those of 2D-ice, the GM mechanism should be valid for 2D-ices. Subsequently only an infinitesimal increase in the band gap 2D-ices as compared to the bulk ices is seen. This was confirmed by our ab-initio simulations.\\
Moreover, h-BN is a wide band gap semiconductor with high thermal  and chemical stability. In both bulk and monolayer h-BN, N atoms and B atoms are hybridized with sp$^2$ at the interlayer forming a honeycomb structure. In bulk h-BN there is weak interactions between each layer of h-BN, such as  electrostatic interactions and Van der Waals interactions, which causes the electrical band gap of both bulk h-BN and 2D-h-BN become  more or less equal~\cite{Today}. Transition from bulk into 2D-TMDs also obeys the same general role, i.e. the band gap of a 2D-TMD is larger than that of bulk.

\subsection{Dielectric constant}
Here, we turn our attention to the main focus of this paper. The dielectric constants extracted from QE represent the combined dielectric constant of 2D-ice and surroundings large vacuum with a height ``$c$". In order to distill the dielectric constant of 2D-ices, we
eliminate the contribution of the vacuum using a capacitance model~\cite{2Dmater}.
In fact, in the out-of-plane (in-plane) direction, the capacitance of the supercell extracted directly from QE code ($\epsilon_{SC}$) is the series (parallel) combination of vacuum capacitance
and the 2D-ice capacitance. This helps us to find the out-of-plane and in-plane relative dielectric constant of 2D-ices using below equations~\cite{2Dmater}:

\begin{figure*}[]
\includegraphics[width=0.45\linewidth]{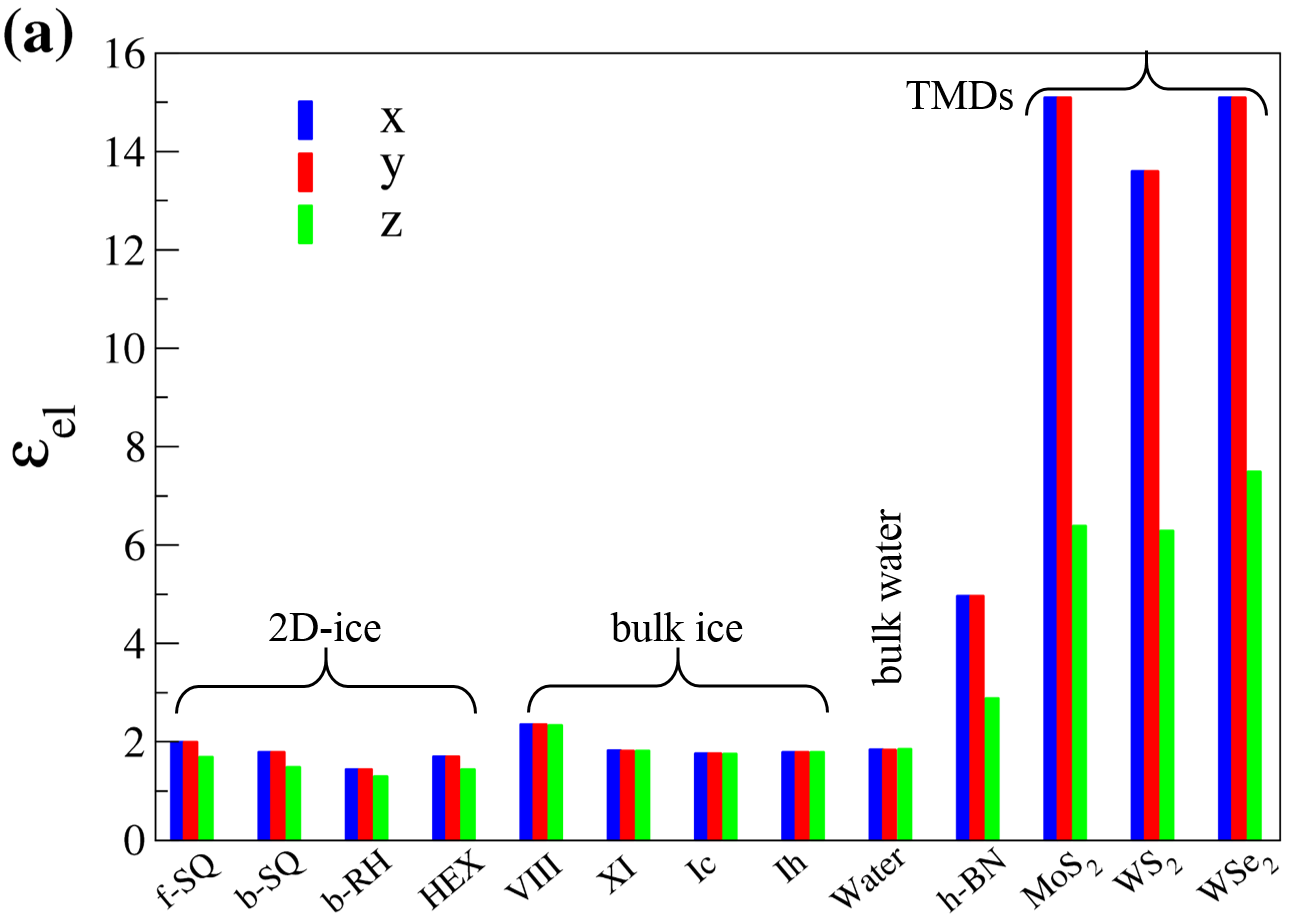}
\includegraphics[width=0.45\linewidth]{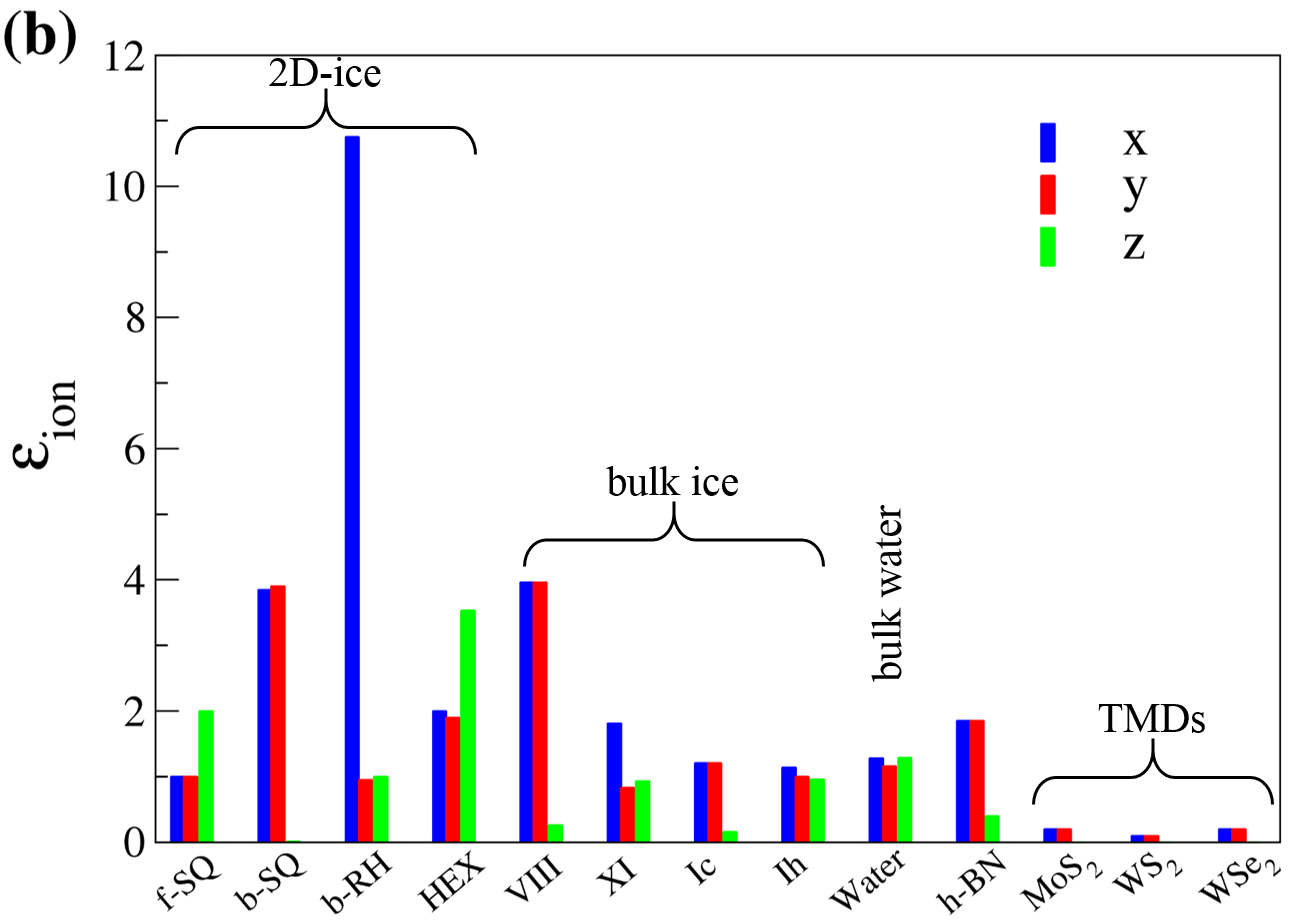}
\caption{ The electronic (a) and ionic (b)  dielectric constant of 2D-ices, bulk ices and TMDs. The data for TMDs were taken
 from Ref.~[14]. }\label{fig4}
\end{figure*}

\begin{figure}[]
\includegraphics[width=0.85\linewidth]{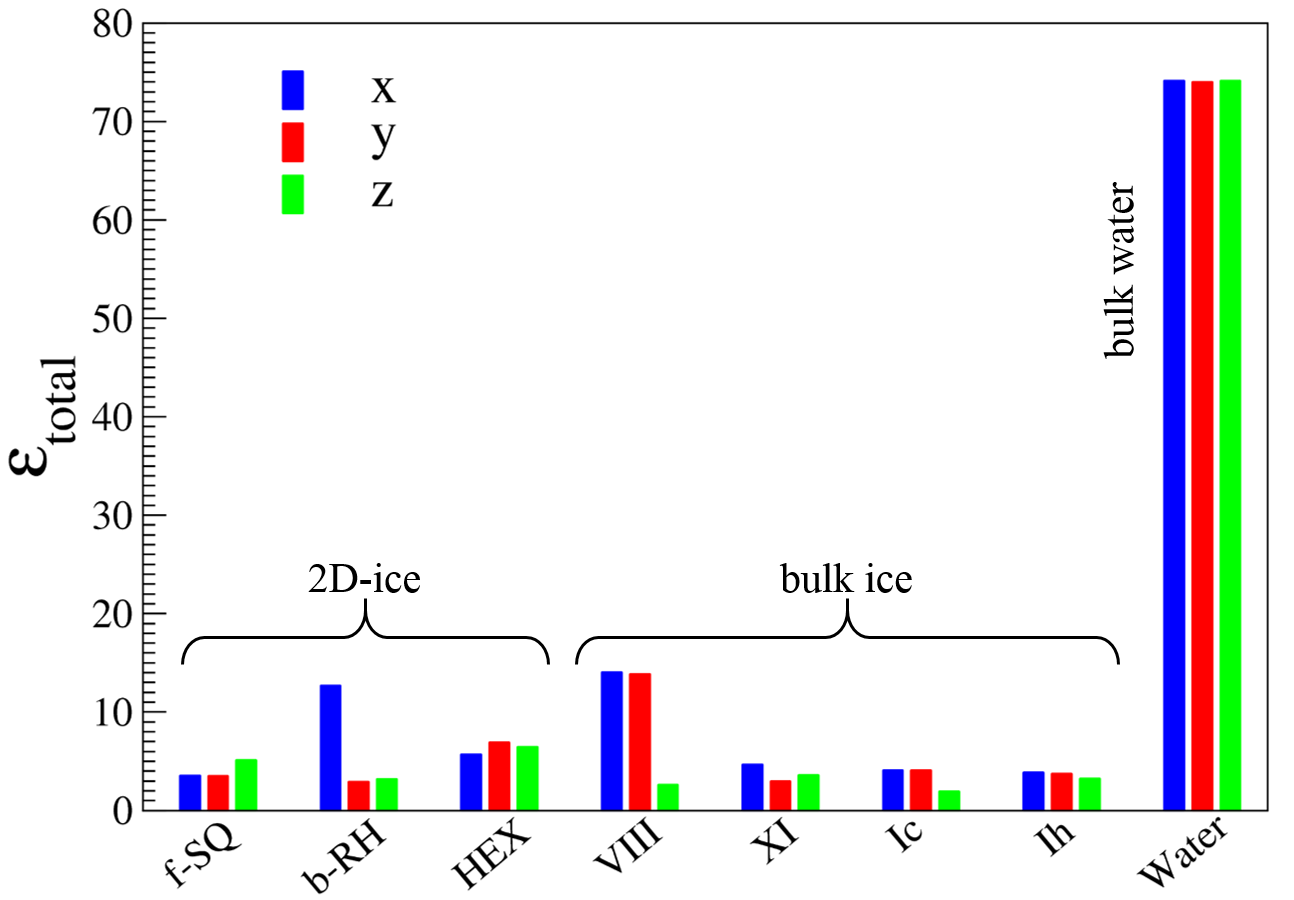}
\caption{The total dielectric constant ($\varepsilon_{total}$) of 2D-ices, bulk ices and bulk water~\cite{Notice}.}
\label{fig5}
\end{figure}

\begin{equation}\label{EqSc}
  \varepsilon^{zz}=[1+\frac{c}{t}(\frac{1}{\varepsilon_{SC}}-1)]^{-1},~
  \varepsilon^{xx,yy}=1+\frac{c}{t}(\varepsilon_{SC}-1),
\end{equation}
``t" is the effective thickness of 2D-ice.

In order to make sure that the used ``c'' values are sufficiently large and the obtained $\varepsilon^{xx,yy,zz}$ were converged well, we performed several additional calculations and found corresponding $\varepsilon_{SC}$. The results show that when ``c'' values change from 12\AA~ to 40\AA,~ the  $\varepsilon^{xx,yy,zz}$ values only increase about 5$\%$. Thus, the used vacuum size  (see Table I)  are sufficiently large.

Using the aforementioned correction, the results for electronic ($\varepsilon_{el}$), ionic ($\varepsilon_{ion}$), and dipolar ($\varepsilon_{dip}$) dielectric constant for 2D-ices are listed in Table I. The main findings are illustrated in Fig.~\ref{fig4} and are listed as:
 \begin{itemize}
   \item   $\varepsilon_{ion}^{xx}\simeq\varepsilon_{ion}^{yy}$ for 2D-ices except for b-RH ice in $\varepsilon_{ion}^{xx}\simeq10 \varepsilon_{ion}^{yy}$, which is due to the remarkable in-plane anisotropic in the  lattice structure of b-RH.

   \item   $\varepsilon_{el}^{xx}\simeq\varepsilon_{el}^{yy}$ for 2D- (bulk) ices and larger (equal to) than $\varepsilon_{el}^{zz}$. The larger the $\varepsilon_{el}^{xx}$ and $\varepsilon_{el}^{yy}$, the more flatness of the structures. This is due to the confined electron clouds in quasi 2D-space.

   \item  $\varepsilon_{el}^{xx}$ and $\varepsilon_{el}^{yy}$  of f-SQ ice are larger than those for other 2D-ices. This might be due to the confining electron in 2D-plane and stronger response of system to an in-plane electric field, where  for other 2D-ices there is a small buckling in their geometry.

%
%
%

   \item Because of the crystalline structure of all the studied crystalline 2D-ices and bulk ices their dipolar contributions are remarkably smaller than that of bulk water.

   \item The electronic dielectric constants of 2D- and bulk ices are smaller than for other 2D-materials such as MoS$_2$, WSe$_2$, and WS$_2$~\cite{2Dmater}. The latter is attributed to the semiconducting nature of the TMDs rather than insulating nature of 2D-ices.
 \end{itemize}

The items  aforementioned are represented in Figs.~\ref{fig6}(a,b).
Consequently the total dielectric constant (Fig.~\ref{fig5}) of the studied ices (except for b-RH and ice VIII) are smaller than 10.

\subsection{Optical dielectric function}
\begin{figure*}[]
\includegraphics[width=0.45\linewidth]{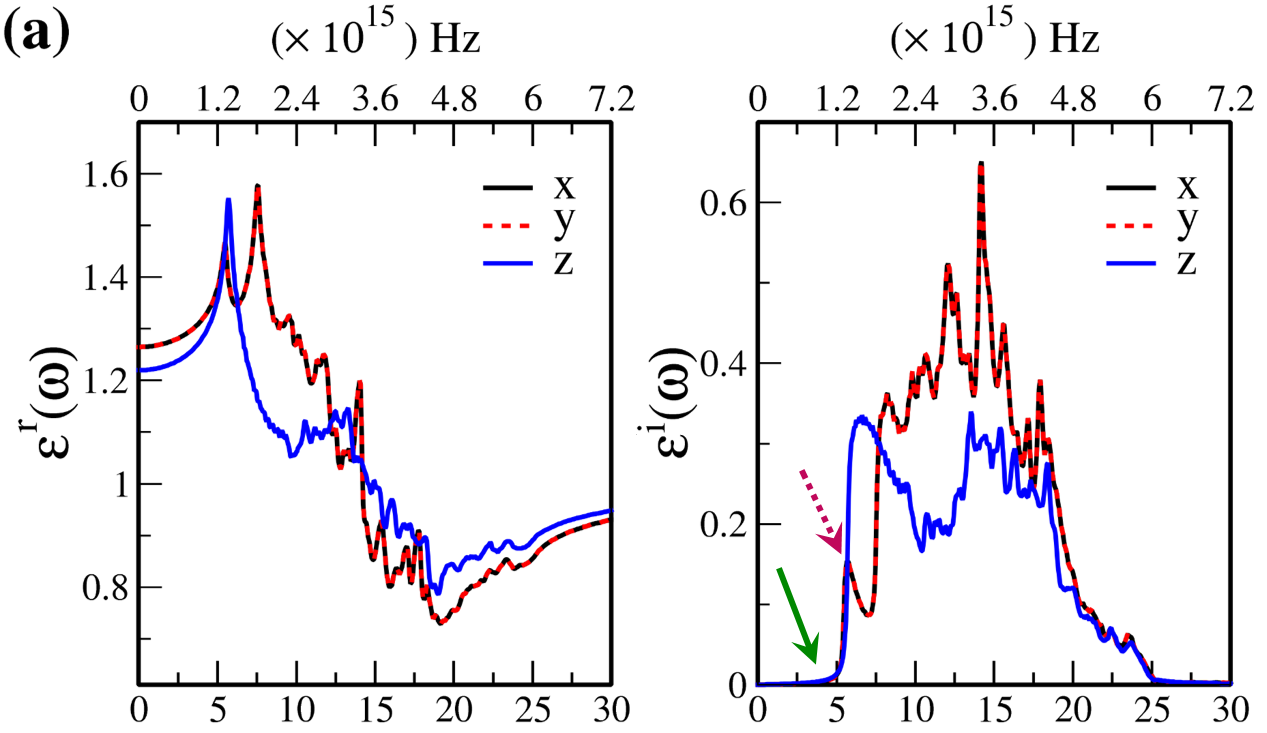}
\includegraphics[width=0.45\linewidth]{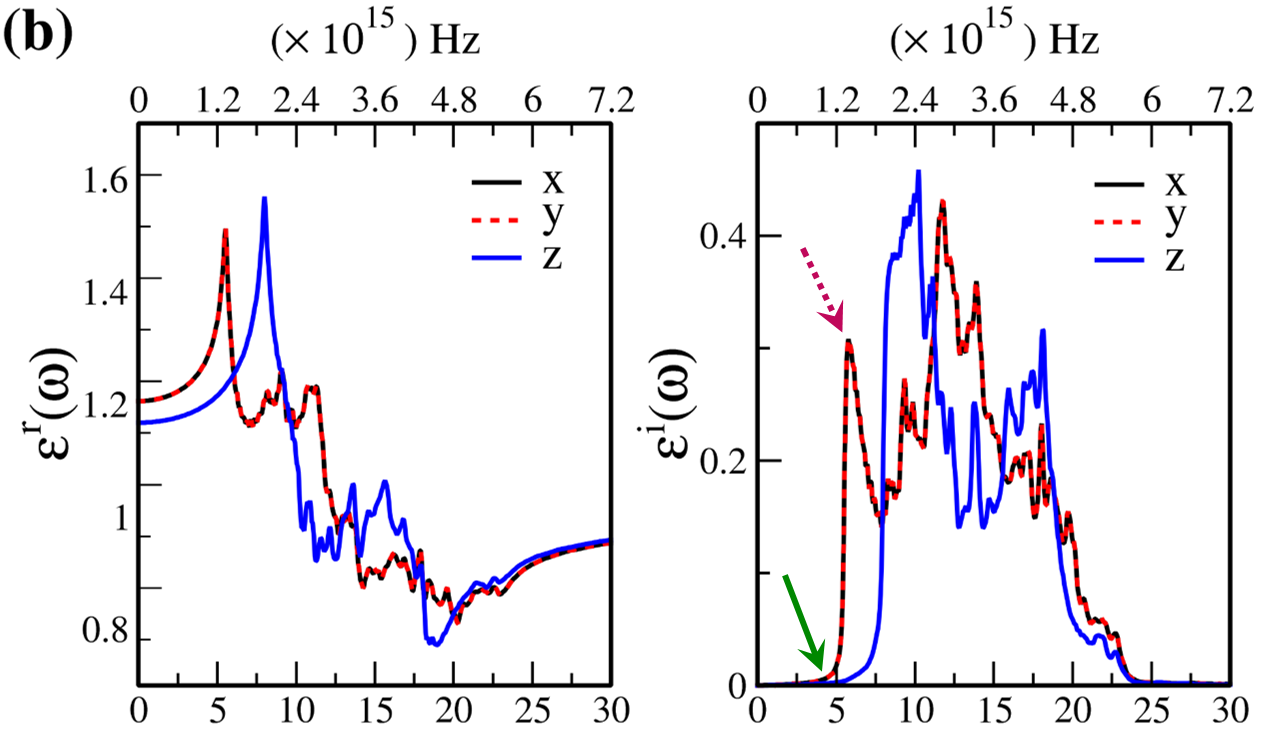}
\includegraphics[width=0.45\linewidth]{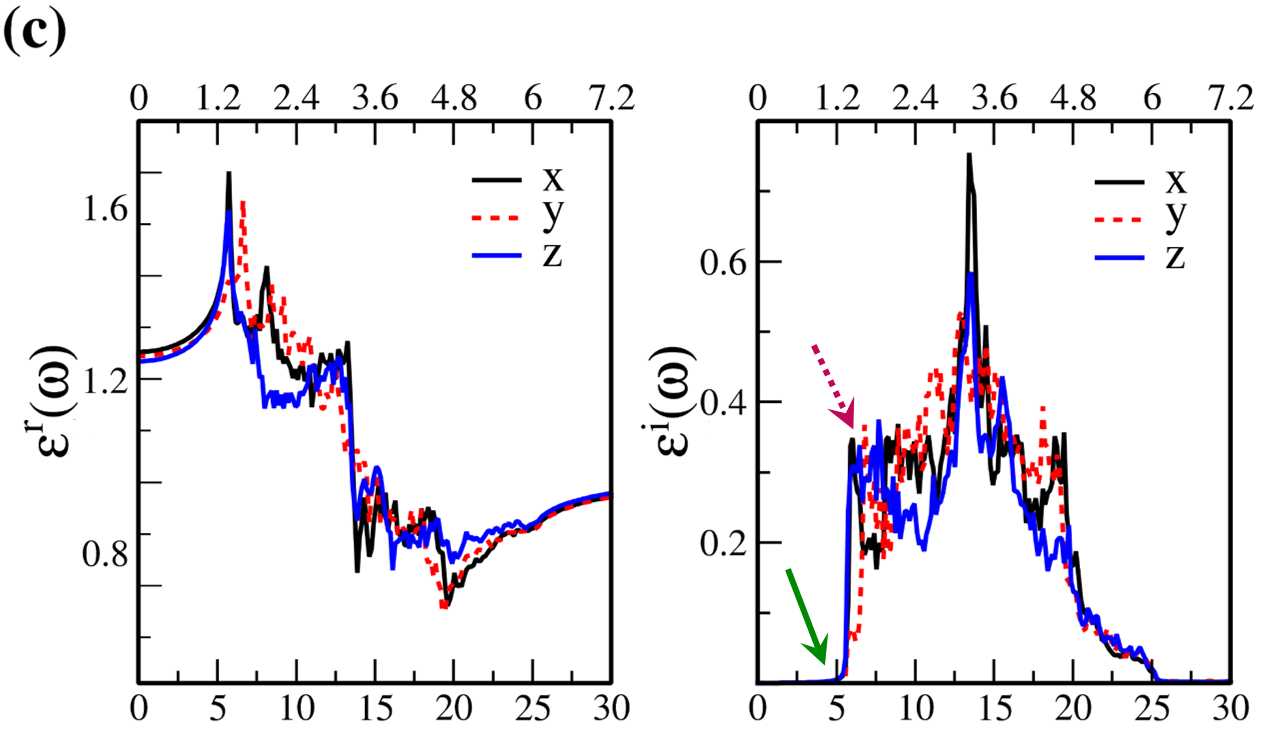}
\includegraphics[width=0.45\linewidth]{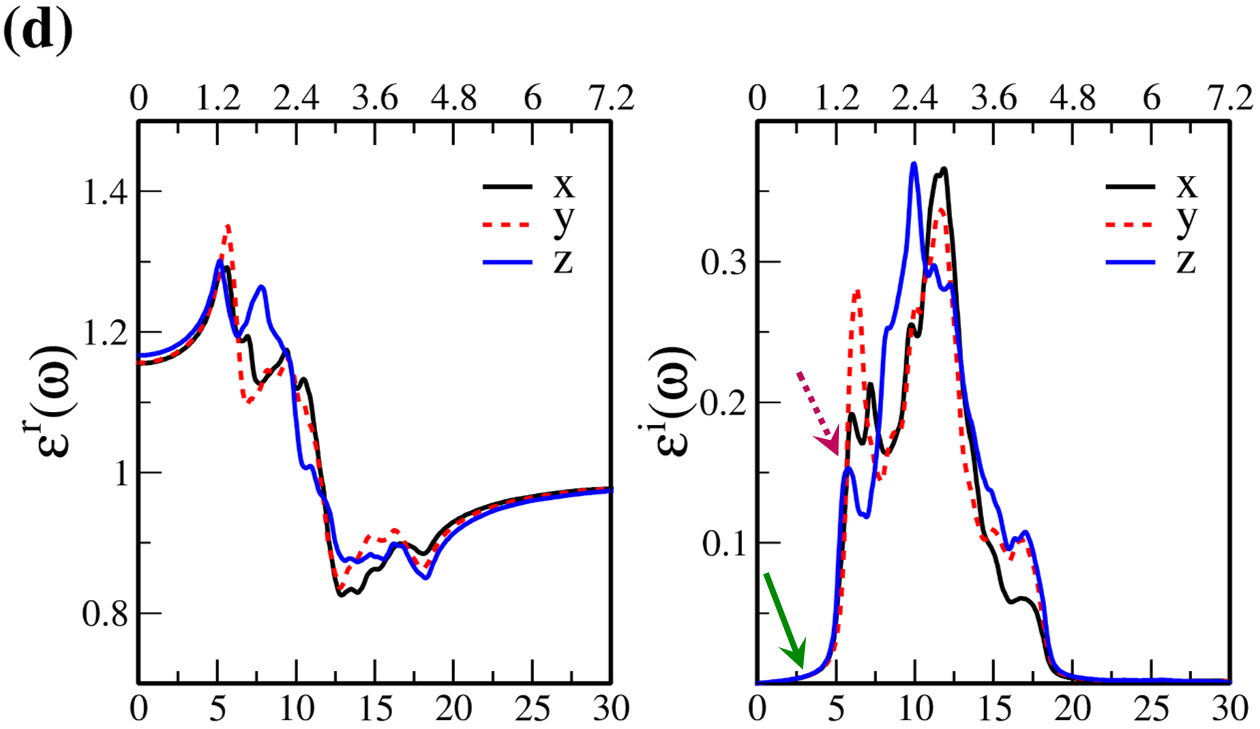}
\includegraphics[width=0.45\linewidth]{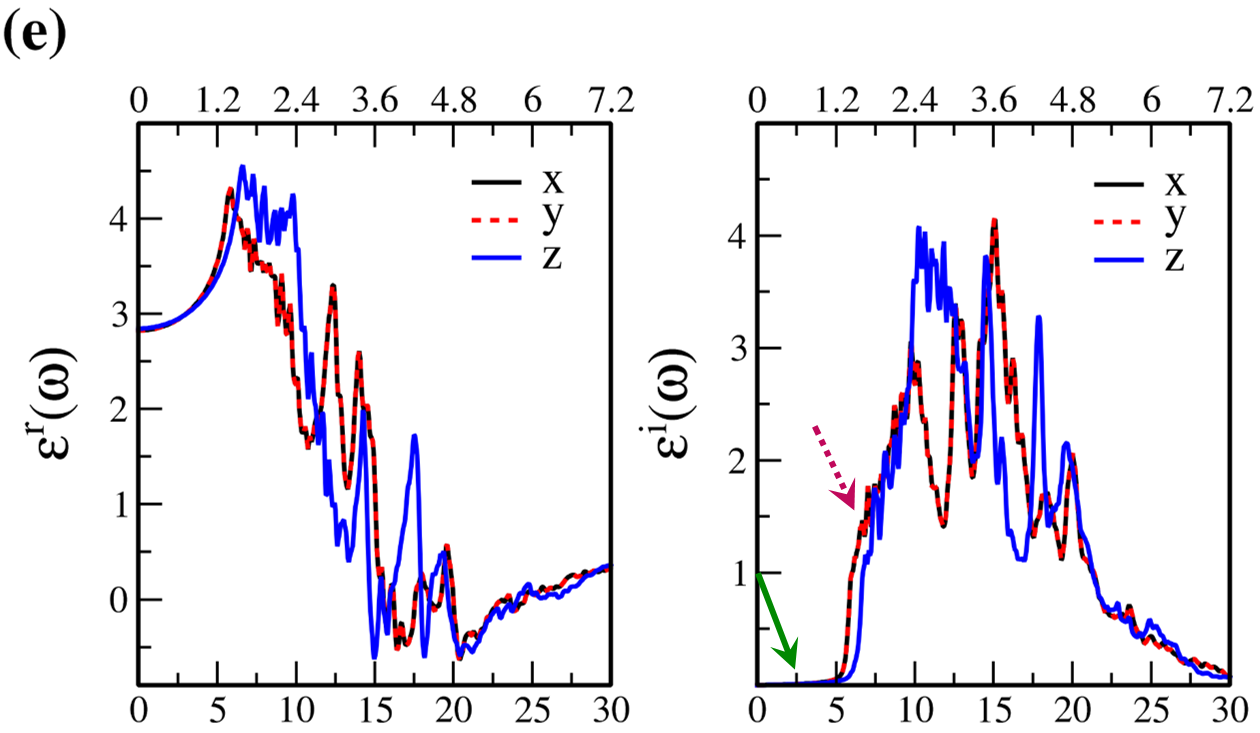}
\includegraphics[width=0.45\linewidth]{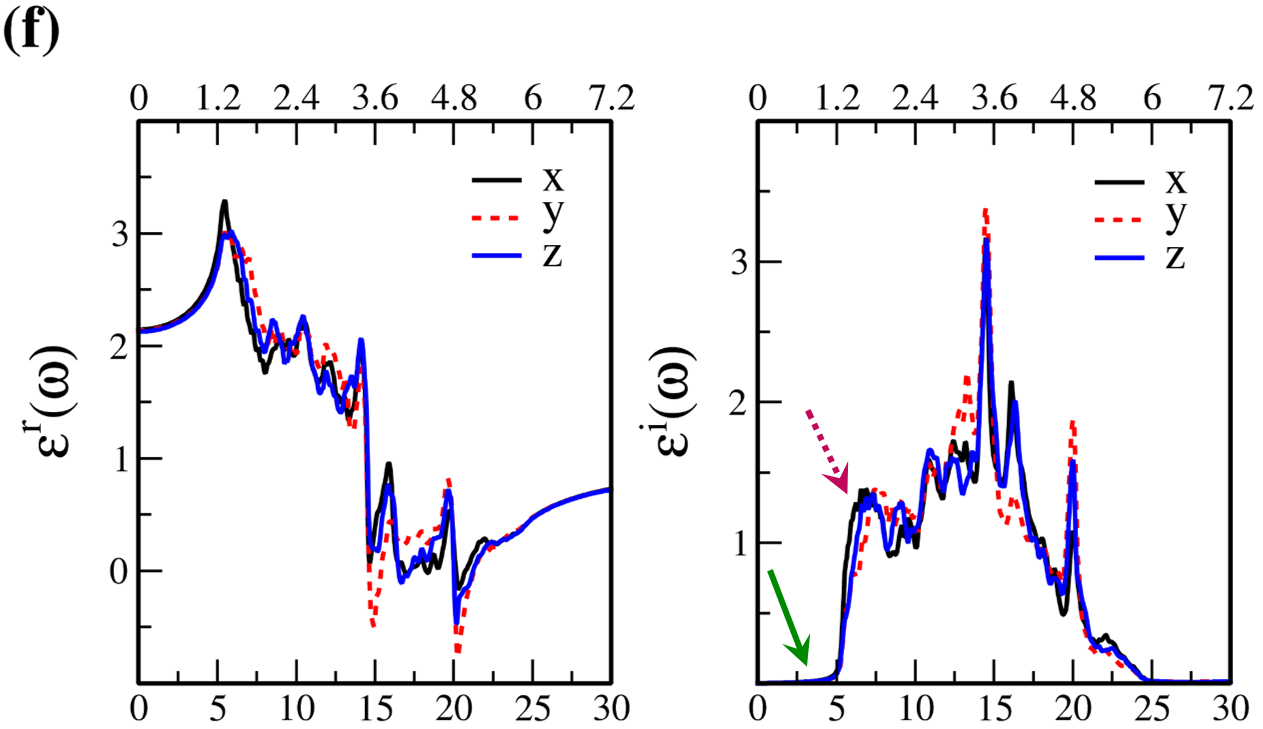}
\includegraphics[width=0.45\linewidth]{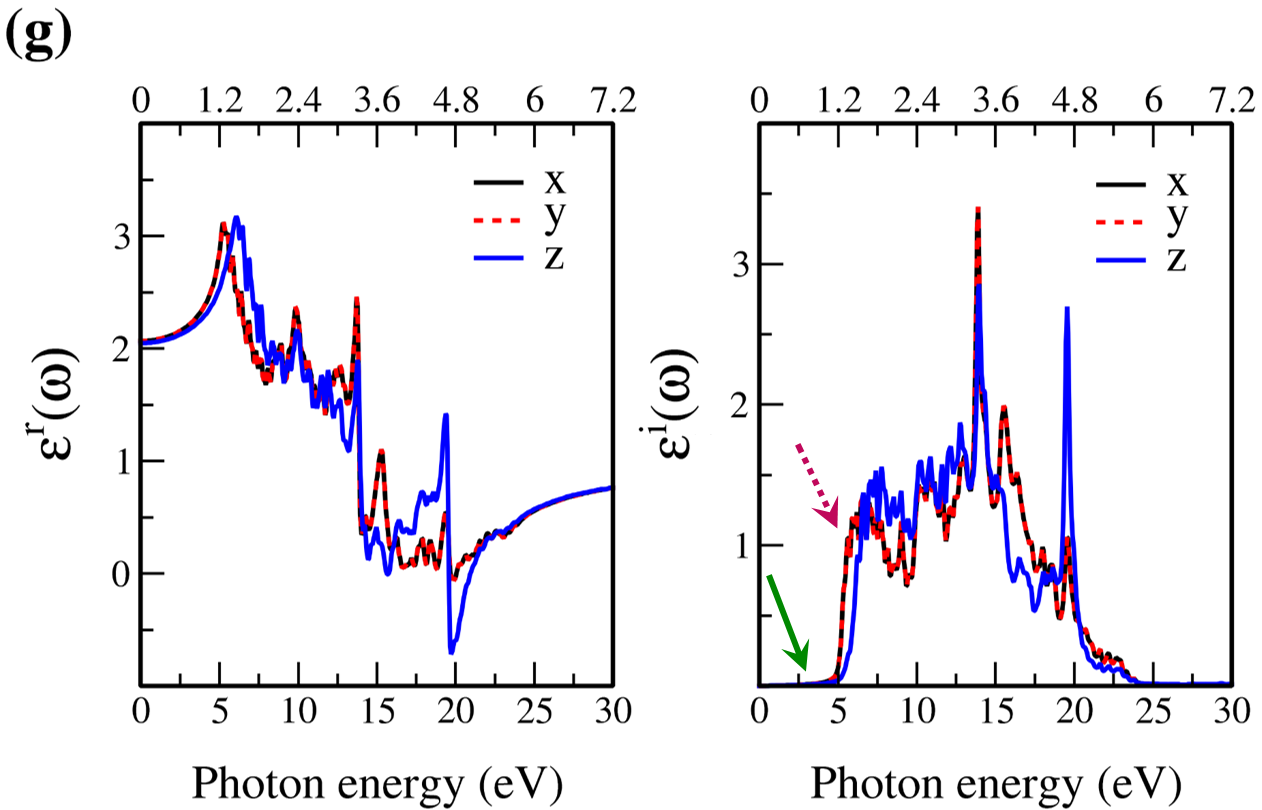}
\includegraphics[width=0.45\linewidth]{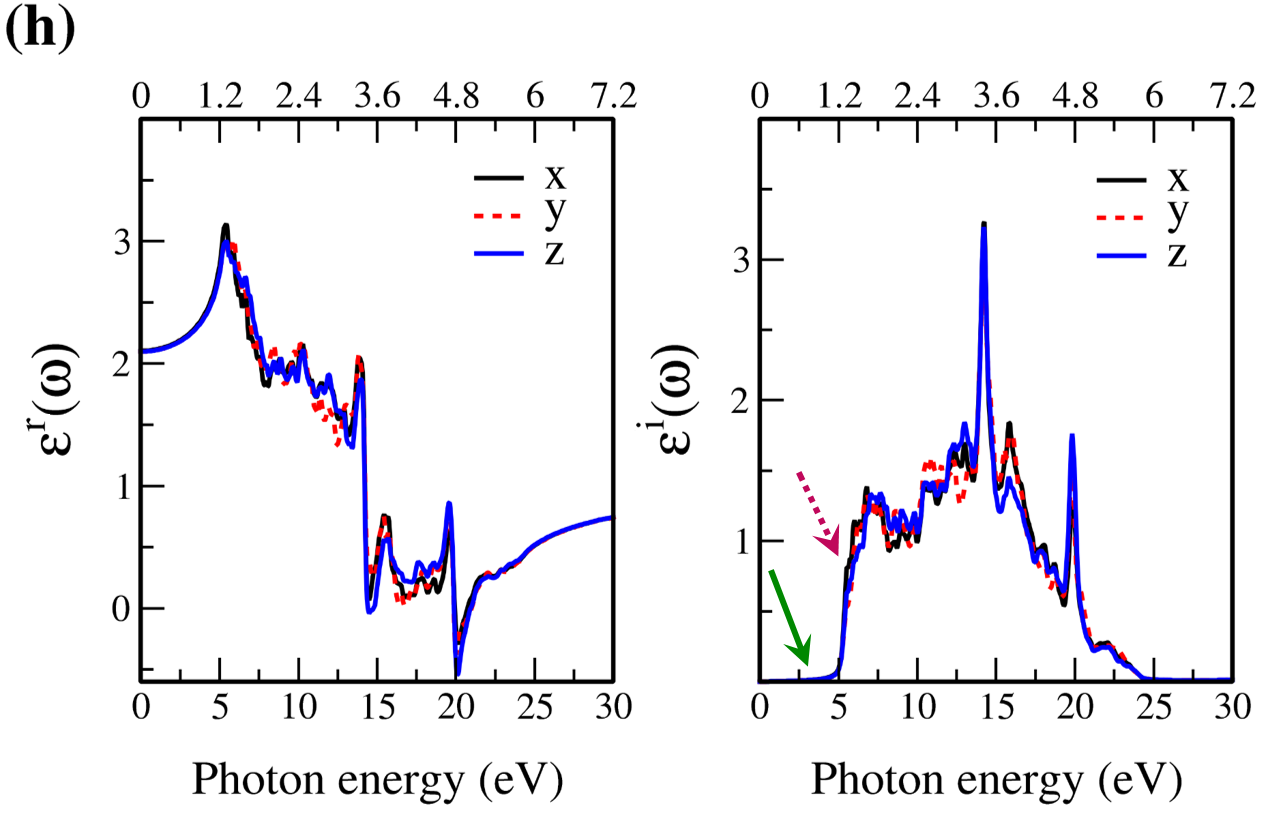}
\includegraphics[width=0.45\linewidth]{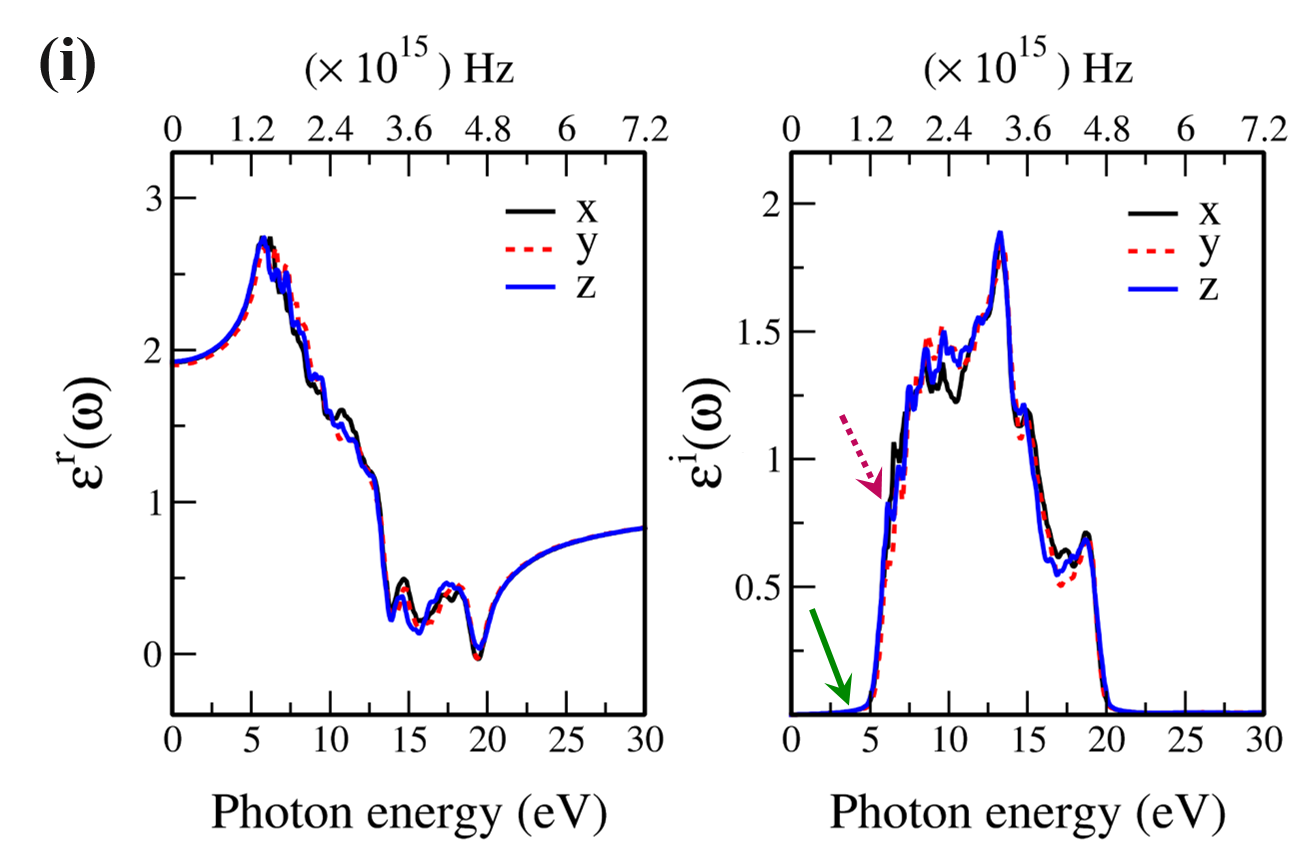}
\caption{The real and imaginary part of dielectric function for
(a) f-SQ, (b) b-SQ, (c) b-RH, (d) HEX, (e) ice VIII, (f) ice XI, (g) ice Ic, (h) ice Ih, (i) bulk water. The green-solid (pink-dashed) arrows refer to
the optical gap (energy gap). }
\label{fig6}
\end{figure*}

 In Fig.~\ref{fig6}, the real ($\varepsilon^r$) and imaginary ($\varepsilon^i$) parts of the optical dielectric function are shown for 2D- and bulk ices  and bulk water. The results show that:

\begin{itemize}
  \item For 2D-ices, $\varepsilon^{xx} \ne \varepsilon^{zz}$ in both real and imaginary parts which is expectable.
  \item Because of isotropic (anisotropic) lattice of the f-SQ and b-SQ ices (b-RH and HEX) for both real and imaginary parts satisfy. $\varepsilon_{el}^{xx}(\omega)=\varepsilon_{el}^{yy}(\omega)  (\varepsilon_{el}^{xx}(\omega)\neq\varepsilon_{el}^{yy}(\omega)$).
  \item  Despite the bulk cubic ice structures (ice VIII and ice Ic), the bulk hexagonal ice structures (ice XI and ice Ih), the dielectric functions are almost equal, i.e. $\varepsilon_{el}^{xx}(\omega)\simeq\varepsilon_{el}^{yy}(\omega)\simeq\varepsilon_{el}^{zz}(\omega)$, for both real and imaginary parts.
\end{itemize}

%
%
%
%
%
%
%
%
%
%

The absorption behavior of ices can be understood by analysing the imaginary part of dielectric function. Also, the absorption edge gives the optical gap ($\Delta_o$). Moreover, the peaks in the $\varepsilon^{i}(\omega)$ function is related to interband transitions in electronic band structure. For instance, the first peak corresponds to the energy gap ($\Delta$). Indeed,   the first critical point calculated in $\varepsilon_{el}^i(\omega)$ is related to the transition from valence band maximum to the conduction band minimum, i.e. the energy band gap.
The green-solid and pink-dashed vertical arrows in Fig.~\ref{fig6} denote the optical gap and energy gap. In general, the optical  gap is equal to the electronic band gap minus the exciton binding energy. In other words, the optical gap and electronic gap should be equal if the many-body perturbation theory is not taken into account. Therefore, our obtained difference between optical gap and electronic band gap is due to the smearing applied in the implemented Kubo-Greenwood approach~\cite{APL2018}. Moreover, note that $\varepsilon^r(\omega=0)$ in Figs. 7 and  $\varepsilon_{el}^0$ listed in Table I, apparently, are different.  However, the difference originates in the two different methods that we employed to calculate them. The first one is the microscopic dielectric function that obtained by using RPA and the second one is the macroscopic dielectric function which was obtained by using DFPT. Thus, they coincide if the same method are used to obtain them. 

\begin{figure*}[]
\includegraphics[width=0.9\linewidth]{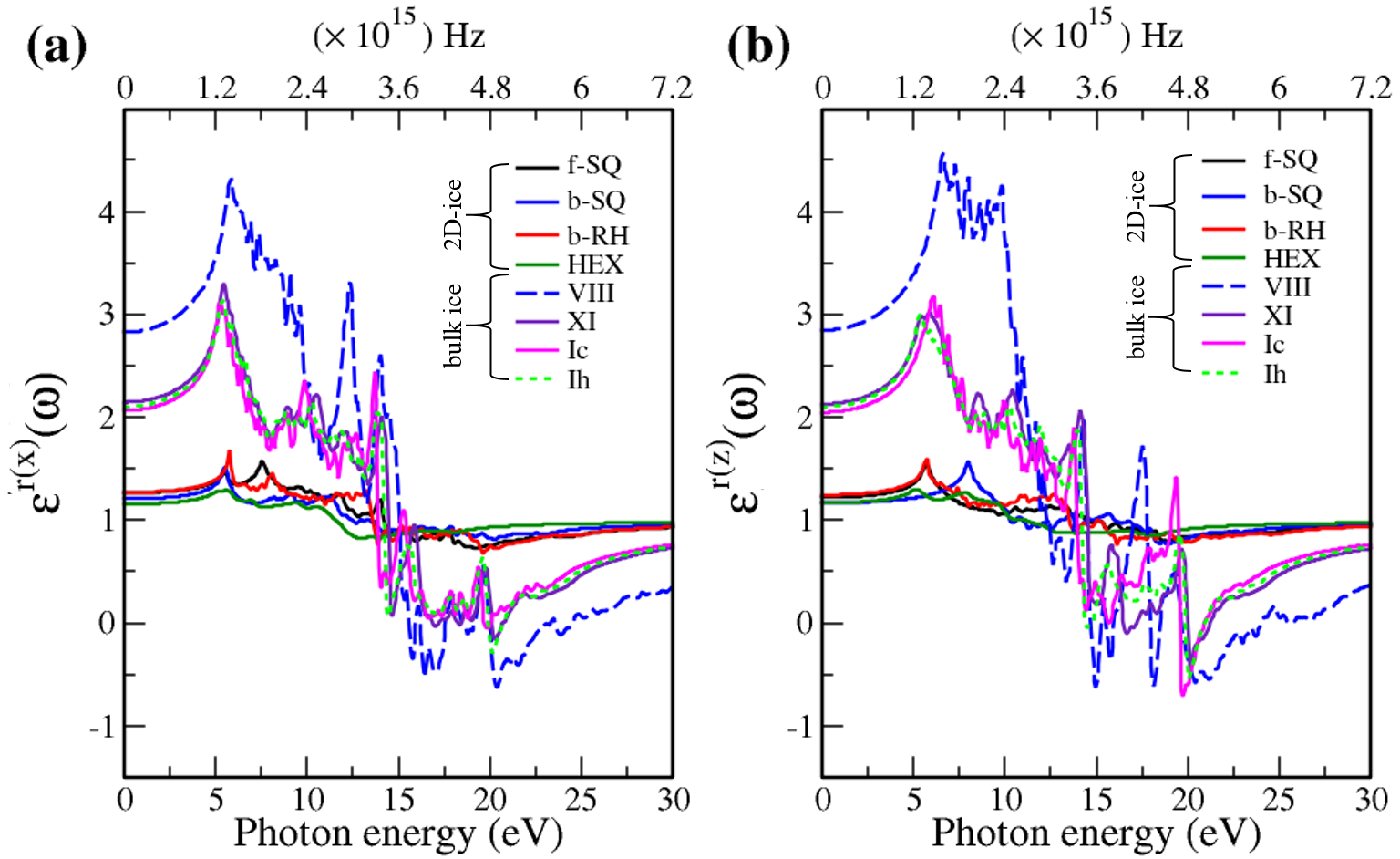}
\caption{The real dielectric function of 2D- and bulk ices: (a) in-plane ($xx,yy$ components) and (b) out-of-plane ($zz$ component).}
\label{fig7}
\end{figure*}

\begin{figure*}[]
\includegraphics[width=0.9\linewidth]{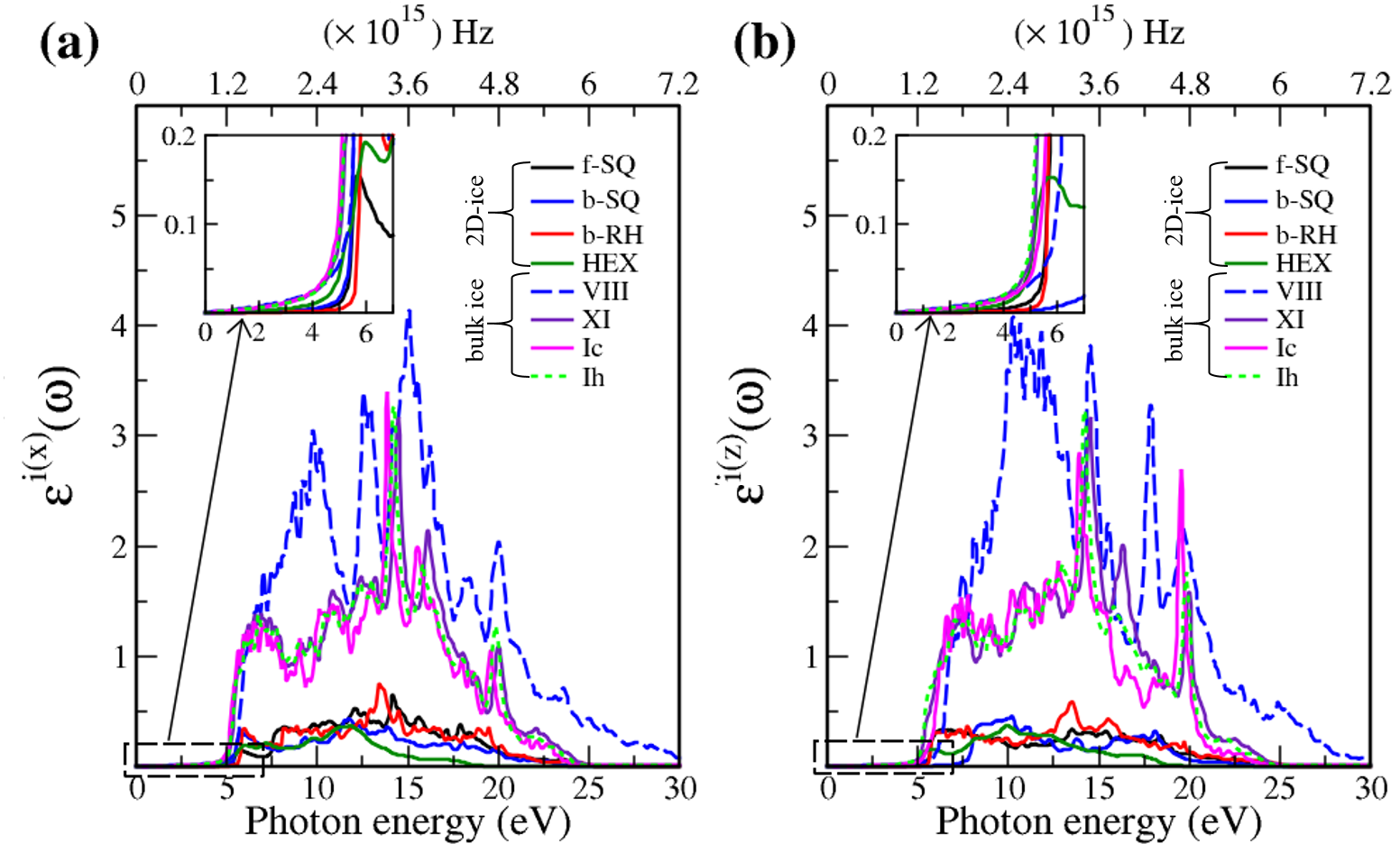}
\caption{The imaginary dielectric function of 2D- and bulk ices: (a) in-plane ($xx,yy$ components) and (b) out-of-plane ($zz$ component).}
\label{fig8}
\end{figure*}

   To compare the optical properties of 2D- and bulk ices, we demonstrated in Fig.~\ref{fig7} (Fig.~\ref{fig8}) the real (imaginary) dielectric functions.
  The results show that:
  \begin{itemize}
  \item The real part of dielectric function has larger values in bulk ices as compared to 2D-ices in low energy region. This is due to the fact that electrons are distributed in 3D space in bulk ice. Moreover, the peaks are larger in 2D-ices as compared to the bulk ices.
  \item The negative values of real dielectric function of bulk ices correspond to the largest electromagnetic wave reflection.
  In fact the inductive properties will dominate in this range of energies with the negative values of real dielectric function. In simple words,  at energy  ranges (e.g. in Ice VIII~the energy range 15-20\,eV  has the negative dielectric function) the electric displacement vector and the electric field vector have opposite directions.

  \item  There is a redshift (move toward lower energy region of the major peaks of $\varepsilon_{el}^i(\omega)$ for bulk ices in comparison to 2D-ices.  In all studied ice, the prominent peaks in  $\varepsilon_{el}^i(\omega)$ correspond to optical transmission which are mainly due to the interband transition from the $p$ valence bands to $s$ conduction bands (i.e. the O-$p$ orbital below Fermi level -valence band- and H-s orbital above Fermi level -conduction band-). PDOS of each system is shown in the RHS panels of Fig. 2.
  
  \item The absorption energy ranges for 2D- and bulk ices are in the ultraviolet spectra ($>3.2 eV$) and visible spectra (between 2 and 3.2 eV), respectively.

  \end{itemize}


 The obtained dielectric functions for ice VIII, ice XI,ice Ic, ice Ih, and bulk water are consistent with the results of Refs.~[32,39], Ref.~[35], and Refs.~[40,41], respectively. Note that $\varepsilon^r(\omega=0)$ in Figs. 7 and  $\varepsilon_{el}^0$ listed in Table I, apparently, are different.  However, the difference originates in the two different methods that we employed to calculate each of them. The first one corresponds to the microscopic dielectric function which was obtained by using RPA and the second one corresponds to the macroscopic dielectric function which was obtained by using DFPT.

\subsection{Mechanical stiffness of 2D-ice}
It is insightful to calculate mechanical stiffness of a typical 2D-ice (the results are selectively presented for f-SQ). We performed an additional calculation for determining the Young's modulus and Poisson's ratio of f-SQ ice. We applied both uniaxial and biaxial strains found the corresponding energies $E_u$ and $E_b$, for uniaxial and biaxial strained systems. Next, the best fits on two equations $E_b=2b u^2_b$ and  $E_u=\frac{1}{2}(b+\mu)u^2_u$, helped us to obtain 2D-bulk modulus ``b" and 2D-shear modulus ``$\mu$". Here $u_u$ and $u_b$ are the applied uniaxial and biaxial strains, respectively.  Consequently, the Young's modulus ($Y_{2d}$) and Poisson's ratio ($\nu_{2d}$) can be computed: ~\cite{zakhar}
\begin{equation}\label{Young}
Y_{2d}=\frac{4b\mu}{b+\mu} ~~and~~ \nu_{2d}=\frac{b-\mu}{b+\mu}
\end{equation}
 The results are $Y_{f-SQ}$=3.6\,N/m and $\nu=0.6$. The Young's modulus of 2D-ice is two orders of magnitude smaller than the one for graphene, i.e. 340\,N/m~\cite{Lee}. This is due to the weak hydrogen bonds in 2D-ices compared to the strong planar covalent bonds in graphene. The corresponding Poisson's ratio of 2D-ice is larger than graphene ($~$0.3). The obtained Young's modulus (Poisson's ratio) for 2D-ice is  more or less equal to bulk modulus of bulk ice. In Table II, we listed Young's modulus/bulk modulus and Poisson's ratio of f-SQ/bulk crystalline ice. To convert N/m unit in $Y_{2d}$ to Pas unit, one may use the simple relation $Y_{3d}=Y_{2d}/t_0$ where $t_0$=3\AA~is the effective thickness of f-SQ ice (see Table I). 

\begin{table}
  \centering
  \caption{The Young's modulus and Poisson's ratio of f-SQ and bulk modulus (B) of three bulk crystalline ice.}\label{table3}
\begin{center}
\begin{tabular}{p{2cm}|p{3cm}|p{2cm}}
\hline
ice& \makebox[3cm][c]{$Y(GPa)$ ($B^*(Gpa))$} & \makebox[2cm][c]{$\nu$} \\[2.5mm]
\hline

f-SQ  &\makebox[3cm][c]{12} &\makebox[2cm][c]{0.6} present work \\[2mm]
\hline
ice VIII~\cite{Klotz} &\makebox[3cm][c]{18$^{*}$}&\makebox[2cm][c]{-} \\[1.5mm]
ice VII~\cite{Klotz} &\makebox[3cm][c]{13$^{*}$} &\makebox[2cm][c]{0.28\cite{Shaw}} \\[1.5mm]
ice Ih~\cite{Klotz} &\makebox[3cm][c]{8.5$^{*}$}  & \makebox[2cm][c]{0.325\cite{Shaw}}\\[1.5mm]
\hline
\hline
\end{tabular}
\end{center}
\end{table}

\section{conclusions}

 In summary, we found that the energy gap in f-SQ, b-SQ, and HEX 2D-ice structures and cubic bulk ices (ice VIII and ice Ic) is direct, whereas b-RH 2D-ice and hexagonal bulk ices (ice XI and ice Ih) have indirect band gap. Underlying lattice structure, symmetry significantly influences the ionic and dipolar terms, but its effects on the electronic dielectric constant are negligible.

 We found the total out-of-plane dielectric constant is larger than 2 for all the studied 2D-ices (except b-SQ) and bulk ices, i.e. $\varepsilon_{total}^{zz}>$2.0 (see Table I).  This clearly shows that the lattice structure of the confined water in recent experiment~\cite{science} is none of the lattice structure of the studied 2D-ices here, and has likely random structure. On the other hand the small out-of-plane dielectric constant of about $\simeq$2.1 (for nanoconfined  water in channels
with heights $h$ $\sim$10\AA)~\cite{science}, should not correspond to a monolayer  water.~Beyond $\sim$15\AA~a nonlinear increase (up to the bulk value) in the dielectric constant was found~\cite{science}.  Therefore, we do not expect to recover experimental data when studying monolayer crystalline ice. It is also interesting to know that there is no reliable experimental data for the in-plane dielectric constant of confined water at sub-nanometer scale slit. Equivalently, the density of trapped water may be much lower than the bulk density~\cite{science}. This motivated us to determine the dielectric properties of amorphous 2D-ice in an ongoing study.\\The optical gap of 2D-ices is found to be larger than that of bulk ices. The absorption energy ranges for 2D- and bulk ices are in the ultraviolet spectra ($>3.2 eV$) and visible spectra (between 2 and 3.2\,eV), respectively. Generally, in bulk materials due to the presence of large number of atoms and merging bunch of adjacent energy levels results in the well-known energy conduction and valance bands. In 2D-materials, due to the smaller number of atoms, the number of energy level decreases giving the narrower energy bands. As a result, energy band gap will increase (the difference between valance band and conduction band).  Also, the larger band gap in 2D-ice will cause a shift of absorption spectrum toward lower wave length (larger energies). In other words, there is redshift in the peaks of $\varepsilon^i$ of bulk ices in comparison to that of  2D-ices~\cite{Yiling}.    \\We believe that our findings not only provide a theoretical background for understanding the different aspects of dielectric properties of confined water, but also gives insights into the light absorption mechanism and corresponding absorption energy range of confined water which might be necessary for further experimental characterizations of 2D-ices.\\
  \\
\section{Acknowledgments}
We acknowledge Nassim Kangarlou for critically reading our paper and giving us useful comments. We would like to acknowledge high-performance computing support from Shahid Rajaee TT-University  sponsored by Iran National Science Foundation (INSF).


\begin{thebibliography}{99}
\bibitem {1810} We found that b-SQ is unstable in our classical molecular dynamics simulation.
\bibitem{Notice}Notice that we substract $\varepsilon_{\infty}$ from $\varepsilon_{total}$ in order to avoid double counting, since $\varepsilon_{\infty}$ appears in both $\varepsilon_{dip}$ and $\varepsilon_{el}$.
\end{thebibliography}
\end{document}